\documentclass[11pt]{article}
\pdfoutput=1

\usepackage{verbatim}
\usepackage{amsmath}     

\usepackage{amssymb}					
\DeclareSymbolFontAlphabet{\amsmathbb}{AMSb}

\usepackage{simplewick} 
\usepackage{mathtools}

\usepackage[titletoc]{appendix}

\usepackage{braket}
\usepackage{cancel}
\usepackage{enumitem}

\usepackage{latexsym}
\usepackage{lscape} 
\usepackage{color}
\usepackage{graphicx}
\usepackage[table]{xcolor}
\usepackage{enumerate}
\usepackage{hhline}
\usepackage{subfig}
\usepackage{multirow}
\usepackage{mathrsfs}
\usepackage{dsfont}
\usepackage{longtable}

\usepackage{amscd}
\usepackage{cite}

\usepackage[colorlinks]{hyperref}
\hypersetup{%
  colorlinks = false,
  linkcolor  = red
}
\usepackage[colorinlistoftodos]{todonotes}

\usepackage{chngcntr}

\counterwithin*{equation}{section}

\allowdisplaybreaks

\usepackage{xspace}

\renewcommand{\d}{\mathrm{d}}

\DeclareMathSymbol{\mg}{\mathrel}{symbols}{"1D}

\newcommand{\ga}{\alpha}
\newcommand{\gb}{\beta}
\newcommand{\gd}{\delta}
\renewcommand{\ge}{\epsilon}

\newcommand{\gl}{\lambda}

\newcommand{\gs}{\sigma}

\newcommand{\gL}{\Lambda}

\newcommand{\Tr}{\mbox{Tr}}

\newcommand{\beq}{\begin{equation}}
\newcommand{\eeq}{\end{equation}}
\newcommand{\barr}{\begin{array}}
	\newcommand{\earr}{\end{array}}
\newcommand{\equ}[1]{\begin{gather} #1 \end{gather}}

\newcommand{\sfrac}[2]{\mbox{$\frac{#1}{#2}$}}
\newcounter{oldcounter}

\newcommand{\qand}{\quad \text{and} \quad}

\newcommand{\U}[1]{\mathrm{U}(#1)}
\newcommand{\SU}[1]{\mathrm{SU}(#1)}

\newcommand{\su}[1]{\mathfrak{su}(#1)}
\newcommand{\uu}[1]{\mathfrak{u}(#1)}

\usepackage{tikz}
\usetikzlibrary{calc}

\newcommand{\diag}{{\text{diag}}}
\newcommand{\Ad}{{\text{Ad}}}

\newcommand{\e}{{\text{e}}}

\newcommand{\order}[1]{\mathcal O\left(#1\right)}

\newcommand{\abs}[1]{\left\vert#1\right\vert} 

\newcommand{\quads}[1]{\quad #1 \quad}

\newcommand{\floor}[1]{\left\lfloor #1 \right\rfloor}

\newcommand{\norm}[1]{\left\lVert#1\right\rVert}

\usepackage{tikz}
\def\checkmark{\tikz\fill[scale=0.4](0,.35) -- (.25,0) -- (1,.7) -- (.25,.15) -- cycle;}

\newcommand{\vev}{\textsc{vev}\xspace}
\newcommand{\cft}{\textsc{cft}\xspace}
\newcommand{\ir}{\textsc{ir}\xspace}
\newcommand{\rg}{\textsc{rg}\xspace}

\newcommand{\upmu}{\mu}
\newcommand{\upupsilon}{\upsilon}
\newcommand{\upomega}{\omega}

\begin{document}

\thispagestyle{empty}

\subsection*{}
\begin{center}
{\Large {\bf A matrix CFT at multiple large charges}
}
\\[0pt]

\bigskip
\bigskip {\large
{\bf Orestis Loukas}\footnote{
E-mail: orestis.loukas@cern.ch},
\bigskip }\\[0pt]
\vspace{0.23cm}
{\it Albert Einstein Center for Fundamental Physics\\
      Institute for Theoretical Physics\\
      University of Bern,\\
      Sidlerstrasse 5, \textsc{ch}-3012 Bern, Switzerland}

\bigskip
\end{center}

\vspace{5cm}
\subsection*{\centering Abstract}
We investigate matrix models in three dimensions
where the global $\SU N$ symmetry acts via the adjoint map.
Analyzing their ground state which is homogeneous in space and can carry either a unique or multiple fixed charges,
we show the existence of at least two distinct fixed points of the renormalization group (\textsc{rg}) flow.
In particular, the one type of those fixed points manifests itself via tractable deviations in the large-charge expansion from the known predictions in the literature.
We demonstrate most of the novel features using mainly the example of the $\SU4$ matrix theory to compute the anomalous dimension of the lowest scalar operator with large global charge(s).

\newpage 
\setcounter{page}{1}
\setcounter{footnote}{0}
{
  \hypersetup{linkcolor=black}
  \tableofcontents
}


\vspace{0.2cm}
\section{Introduction}

Conformal field theories (\textsc{cft}s) are of great importance in modern physics.
They appear at the fixed points of the
\textsc{rg} flow in a variety of different systems, ranging from critical phenomena to quantum gravity and string theory.
Unfortunately, most of those \textsc{cft}s lack nice perturbative limits making any analytic investigation harder or impossible.

%
%
However, as it was first observed in \cite{Hellerman:2015nra}, there exist certain strongly coupled \textsc{cft}s in the infrared (\textsc{ir}) in $2+1$ dimensions with some global symmetry for which a Wilsonian effective action can be written down in a meaningful way. 
In fact, those \textsc{cft}s are found to be effectively at weak coupling by considering sectors of the theory at fixed and large values of the associated global charge $Q$. Recall that $Q$ is dimensionless in natural units and $1/Q$ becomes the controlling parameter of a perturbation series (in a spirit similar to large spin theories \cite{Komargodski:2012ek,Alday:2016njk,Kaviraj:2015cxa,Dey:2017fab}) around a non-trivial vacuum --being different from the vacuum of the full theory-- which fixes the charge in the given sector. 
%
Small fluctuations around this vacuum  are parametrized by Goldstone fields with non-Lorentz invariant dispersion relations\footnote{Such systems at finite charge density have been studied previously in the literature, see e.g.\ \cite{Nicolis:2011pv,Watanabe:2013uya}. As the charge density was not taken large, though, the outlined perturbative character did not manifest itself in those older studies.}, which appear
as a result of 
breaking the internal global symmetry group together with  conformal invariance. Any higher corrections are suppressed by appropriate powers of  $1/Q$.
This allows the perturbative  computation of anomalous dimensions and fusion coefficients in the three-dimensional \cft in the regime (with $\gL$ the \textsc{ir} cut off and $\mathcal V$ the volume of the two-sphere) where
\equ{
	\label{Intro:ScaleHierarchies}
	\sqrt{\frac{1}{\mathcal V}} \,\ll \, \gL \, \ll \, \sqrt{\frac{Q}{\mathcal V}	}
	~.
} 

This rather unexpected outcome was 
confirmed \cite{Banerjee:2017fcx} via Monte-Carlo simulations of the $O(2)$ model on the lattice.
At the analytic level, the large-charge construction was
verified and systematized\footnote{An independent derivation of large-charge theory in terms of conformal bootstrap has been recently given in \cite{Jafferis:2017zna}. Moreover, Large-$R$ expansions in models with $\mathcal N=2,4$ superconformal symmetry  have been mainly considered in \cite{Hellerman:2017veg,Hellerman:2017sur}.} in \cite{Alvarez-Gaume:2016vff} using the paradigm of $O(n)$ vector models (with the field content in the vector representation of the global symmetry group).
%
%
Differently from the situation in chiral symmetry breaking where the low-energy spectrum is dynamically determined, 
various non-trivial symmetry-breaking patterns can appear in the sectors of a theory at fixed and large charge.
{Instead of starting from a concrete symmetry breaking pattern in the effective description (see \cite{Monin:2016jmo} for this approach), we shall use the linear sigma model as an intermediate tool to find the light  spectrum (gapless modes) relevant for the low-energy physics, in the spirit of \cite{Alvarez-Gaume:2016vff}.
In more detail, the procedure established there} to analyze such large-charge sectors of the 
\textsc{cft} at hand is the following:
\begin{itemize}[leftmargin=0.7cm]
	\setlength\itemsep{0.2em}
	\item 
	{Assume an order parameter for the critical theory}
	\item
	Specify the global symmetry group and how it acts on the order parameter 
	\item
	Write a Wilsonian effective action in the \textsc{ir} which enjoys all the global and local symmetries
	\item
	Use this action to solve the classical problem of fixing the charge and establish the vacuum
	\item
	Deduce the 
	light spectrum relevant for the low-energy physics in the large-charge sector by quantizing the fluctuations on top of the previously determined classical ground state
	\item
	Ensure the stability of the expansion under quantum corrections by integrating out  heavy modes, {thus verifying the self-consistency of the effective description}
\end{itemize}
In \cite{Loukas:2017lof} the large-charge techniques were extended in a similar spirit with the aim of understanding strongly coupled $\SU N$ matrix models (with the field in the adjoint representation).
As a working example for that, the scalar $\SU3$ matrix theory was examined, which is of phenomenological interest due to its relation to the $\mathbb{CP}^2$ universality class \cite{Delfino:2015gba}. It turns out that the $\SU3$ matrix model flows in the \textsc{ir}
to a fixed point which produces the same qualitative predictions to  leading orders in the large-charge expansion as the  Wilson-Fisher fixed point of the vector model theory. 

%
%
In this paper, we put the latter finding in perspective and provide fixed-charge  solutions for matrix models with larger  $\SU N$ symmetry groups.
Specifically, we find (at least) two 
fixed points of the \textsc{rg} flow which produce distinct predictions in the large-charge expansion to tractable order.
In the first class, the low-energy theory mimics the structure of the Wilson-Fisher fixed point. In particular, the vacuum configuration with the lowest possible energy (which is homogeneous in space) allows us to fix only one independent charge scale, i.e.\ there can only exist one independent $\U1$ charge $Q$, which is non-zero (and large). Trying to fix an additional 
$\U1$ scale at this fixed point will inevitably  introduce inhomogeneities in space, as it was observed in \cite{Hellerman:2017efx} for a similar setup. 
Contrary to that, the second class of fixed points in the $\SU{N}$ matrix theory 
allows us to independently fix up to $\floor{N/2}$ different charge scales $Q_j$ in the low-energy effective description, while {the ground state still remains} 
homogeneous in space. 
Obviously, at least one of those independent $\U1$ charges needs to be taken large 
for the perturbative analysis to apply in the sense of Eq.\ \eqref{Intro:ScaleHierarchies}. 

%
%
We exemplify these qualitative and quantitative differences by computing the anomalous dimension $\Delta$ of the lowest scalar operator\footnote{Operators with large charge and non-zero spin have been also recently studied in \cite{Cuomo:2017vzg}.} with a particular charge configuration in the three-dimensional flat-space 
\cft that describes the matrix model in the \ir. (By the state-operator correspondence this $\Delta$ is mapped  on the cylinder to the lowest-energy state characterized by the same charge assignment.)
Concretely, we take the example of the $\SU4$ matrix model which possesses the smallest global symmetry group exhibiting novel features. 
A scalar operator either with charge $Q$ at the former fixed point ($P=1$) or with charges $Q_1=Q$ and $Q_2$ at the latter fixed point ($P=2$)
has an anomalous dimension that can be formally organized as an asymptotic expansion in $1/Q\ll1$:
\equ{
	\label{Intro:SchematicEnergyExpansion}
	\Delta^{(P)} 
	= \ga_P \,{Q}^{3/2}
	+
	\beta_P  \,{Q}^{1/2}
	-0.0937
	-f_P
	+\mathcal O (Q^{-1/2})
	\quads{\text{with}} P=1,2 
} 
and $\Delta^{(1)}\equiv\Delta(Q)$ whereas $\Delta^{(2)}\equiv\Delta(Q,Q_2)$ and $\ga_2\equiv\ga(Q_2/Q)$, $\beta_2\equiv\beta(Q_2/Q)$.

In the leading condensate part of this formula $\ga_P$ and $\beta_P$ are ignorance coefficients of order one which have to be determined via non-perturbative methods.
Quantitatively, one expects them to be different at different fixed points $P$ of the \textsc{rg} flow.
Already to leading order in $Q$, we shall see a clear difference between the two fixed points. The ignorance parameters $\ga_1$ and $\beta_1$ at the former fixed point depend only on the microscopic details of the underlying physical system, but not on the global charge $Q$ we select.
On the other side, 
the ignorance coefficients at the latter fixed point  depend on the (generically order-one) ratio $Q_2/Q_1$ of the two $\U1$ scales  we choose. Hence, $\ga_2$ and $\gb_2$ are expected to be different for different values of $Q_2/Q_1$ and have to be determined via non-perturbative methods for each fixed ratio, separately.
As it is beyond the analytic scope of the current paper, 
we leave this as an open question for a non-perturbative treatment of the theory.

%
%
Contrary to the leading ignorance parameters, at order one at the perturbative level there is a universal -- independent of the fixed point -- contribution.
Most crucially, though, a qualitative difference appears at order one: $f_P$ represents an order-one contribution which depends on the class of fixed points we look at. If the lowest operator carries 
one non-vanishing 
$\U1$ scale $Q$,
then $f_1=0$. In the case when 
$Q_1$ and $Q_2$ are simultaneously non-zero, then $f_2$ poses a non-vanishing contribution, which depends on 
the microscopic details of the 
physical system as well as on the charge ratio $Q_2/Q_1$.
%
%
The main objective of this work is to see how $f_P$ appears and to justify the related asymptotic expansions for $\Delta^{(P)}$ by studying their behavior under various charge configurations.

\subsubsection*{Overview of the paper}

To this end, in Section~\ref{sc:LSM_Theory} we lay out the matrix model we wish to investigate and review the necessary theoretical framework to perform our large-charge analysis.
Subsequently, we separately consider the two classes of fixed points. In Section~\ref{sc:TwoMus} we look in great detail at the 
novel fixed point ($P=2$), while in Section~\ref{sc:OneMu} the more familiar situation ($P=1$) is discussed which is similar to Wilson-Fisher with at most one non-vanishing $\U1$ charge scale. 
The analysis is done using the concrete example of the $\SU4$ matrix model exhibiting a sufficiently large symmetry group to accommodate both classes of distinct fixed points. We also outline how the generic $\SU N$ theory works.
In both sections we derive expressions of the form \eqref{Intro:SchematicEnergyExpansion} for the vacuum energy of the homogeneous charged state (equivalently the anomalous dimension of the lowest scalar operator), which we compare and contrast.
Ultimately, in Section~\ref{sc:NonLinear} we  provide  the effective actions (using the non-linear sigma model description) for some of the non-trivial light spectra we have derived. 
In Appendix \ref{app:Propagators}, expressions for various propagators used throughout the derivation are given.

\section{The linear sigma model}
\label{sc:LSM_Theory}

To study the behavior of a particular \textsc{cft} at large charge it is not enough to look at the global symmetry, we also need to specify how this symmetry acts on the order parameter (i.e.\ the matter content) of the critical theory.
In the current paper we choose to work with matter in the adjoint representation of the global $\SU N$ group, meaning
our order parameter is a traceless hermitian matrix, $\Phi \in \su N$. 
In this section we review how to write
a linear sigma model in $\Phi$  
and introduce the necessary notation and techniques to be implemented in Sections~\ref{sc:TwoMus} and \ref{sc:OneMu}.

\subsection{Classical analysis and the fixed point structure}

First, we set up the classical problem at finite charge(s) within the framework of the linear sigma model and comment on the qualitatively distinct fixed points that emerge, already by considering the classical theory.

\paragraph{The Lagrangian formulation.}
Our starting point is a Wilsonian action in the \textsc{ir} for the field $\Phi$ living in $\mathbb R\times\Sigma$ (where $\Sigma$ can be any well-behaved, compact two-dimensional manifold), 
\begin{equation}
\label{LSM:FullAction}
S= \int\d t \d \Sigma\, \mathcal{L} =\int \d t \d \Sigma \left[\sfrac{1}{2} \Tr\, (\partial_\mu \Phi \partial^\mu \Phi)  - V(\Phi)\right]
~,
\end{equation}
in terms of a scalar potential which we choose as (we comment below on possible generalizations in relation with the fixed-point structure)
\begin{equation}
\label{LSM:ScalarPotential}
V(\Phi) = \frac{\mathcal R}{16}\,  \Tr\,\Phi^2 + g_1 \Tr\, \Phi^6 + g_2 \left(\Tr\,\Phi^2\right)^3  + g_3 \left(\Tr\,\Phi^3\right)^2 + g_4 \Tr\,\Phi^4\,\Tr\,\Phi^2 ~.
\end{equation}
$\mathcal R$ is the scalar curvature of $\Sigma$ and $g_i$ for $i=1,2,3,4$ are dimensionless Wilsonian couplings of order one. 
A necessary condition for the model under consideration to make sense in the first place,  is that the scalar potential \eqref{LSM:ScalarPotential} is stable.
In detail, the potential has to be bounded from below, meaning  it cannot have a runway behavior  at infinity, when $\Tr\,\Phi^2\rightarrow\infty$. This amounts to a set of conditions for the couplings $g_i$.
Only inside the cone defined by this set of conditions in the space spanned by $\lbrace g_i\rbrace$ our analysis is valid.
Since the action under consideration is a tool we are using to derive the low-energy \textsc{dof}s, the precise form of the cone is not of particular interest.
We are content to know that there exists at least a non-trivial region inside the cone. 
For instance, take all $g_i\geq0$, then obviously $V(\Phi)$ is well bounded from below.
By trace cyclicity we readily see that the action is invariant under global $\SU N$ transformations acting on the order parameter via the adjoint map, 
\equ{
	V \in \SU N\,: \quad\Phi ~\rightarrow~ \Ad[V] \Phi := V \Phi V^{-1}
	~.
}
To this global symmetry transformation there exists an associated Noether current
\equ{
	\label{LSM:NoetherMatrix}
	J_\mu = i \left[\Phi,\partial_\mu \Phi\right]
	~.
}
Assigning to the field operator the (naive) classical mass dimension $[\Phi]=1/2$, the action under consideration becomes also scale invariant.

We will use this action to find the symmetry-breaking pattern associated to fixing some large scale $Q \gg1$ in the infrared \textsc{cft}.
%
%
The light spectrum (i.e.\ gapless modes) arising due to the derived symmetry breaking comprises 
the good low-energy \textsc{dof}s that are used in Section~\ref{sc:NonLinear} to write down the 
appropriate non-linear sigma model for a given large-charge configuration. 
Therefore, it is sufficient for our purposes to look at the particular linear sigma model described by Eq.~\eqref{LSM:FullAction} to 
deduce the relevant Goldstone spectrum. 
In addition, the 
specified action is able to capture all the physics in the large-charge expansion 
up to order one, which 
{can be more intuitively understood by looking at the gravity dual \cite{Loukas:2018zjh} of the investigated matrix theory.}

Incidentally, the kinetic and curvature terms of the Lagrangian described by Eq.\ \eqref{LSM:FullAction} are invariant under $O(N^2-1)$ global transformations.
Once the parameters in the potential $V(\Phi)$ of the linear sigma model are adjusted such that $g_1=g_3=g_4=0$, the full action of our matrix model enjoys the enhanced $O(N^2-1)$ symmetry. 
In such a coupling configuration we recover the vector model paradigm, albeit in a different parametrization.
Since the vector model theory has been already explored in \cite{Alvarez-Gaume:2016vff}, we do not discuss it in this paper, i.e.\ we always take at least $g_3\neq0$.

\paragraph{The Hamiltonian formulation.}
Since $\Phi$ is hermitian, we can diagonalize it, 
\equ{
	\label{PhiEigendecomposition}
	\Phi = U A U^\dagger
	\quads{\text{with}}
	U \in \SU N/\U1^{N-1}
}
and  obtain the eigenvalue matrix
\equ{
	\label{LSM:EigenvalueMatrix}
	A=\diag\left(a_1,...,a_N\right)
	\quads{\text{with}} a_1 +...+a_N=0
	~.
}
This eigendecomposition makes plausible to define the angular velocity 
together with the canonically associated angular momentum matrix
\equ{
	\label{AngularMomentumInTheBody}
	\omega = -i U^\dagger \dot U
	\qand
	K =\left(\frac{\partial \mathcal L}{\partial\omega}\right)^T = U^\dagger J_0 U = \left[\left[\omega,A\right],A\right]
	~.
}
Using these definitions 
we can compactly write the Hamiltonian corresponding to Lagrangian \eqref{LSM:FullAction} as
\equ{
	\label{LSM:FullHamiltonian}
	\mathcal H = \sfrac12 \Tr \left( \pi_A^2 + (\nabla A)^2 + \left[U^\dagger \nabla U, A\right]^2 \right) + \sfrac{1}{2} \sum_{i\neq j} \frac{\vert K_{ij} \vert^2}{(a_i-a_j)^2} + V(A)
	~,
}
with the conjugate momentum to $A$ being $\pi_A^T = \partial\mathcal L/\partial \dot A$. Notice that 
the kinetic part of the Hamiltonian is written as a sum of squares. 
In this work we are interested in the lowest energy configuration at finite charge density $J_0$. From Eq.\ \eqref{AngularMomentumInTheBody} we see that $J_0\neq0$ implies $K\neq0$. By inspecting the classical Hamiltonian it follows that the charged state  of lowest energy is described by a static ($\dot A=0$) and homogeneous in space ($\nabla A=0$ and $\nabla U=0$) 
solution to the Euler-Lagrange \textsc{eom}s,
\equ{
	\label{LSM:EOMs}
	\ddot \Phi_\text{cl} = -V'(\Phi_\text{cl})
	~.
}
Any vacuum $\braket{\Phi}$ in this paper will be of the form $\Phi_\text{cl}\equiv\Phi(t)$. 
In a static and homogeneous regime the classical Hamiltonian \eqref{LSM:FullHamiltonian} simply becomes
\equ{
	\label{LSM:HomogeneousHamiltonian}
	\mathcal H = \sfrac{1}{2} \sum_{i\neq j} \frac{\vert K_{ij} \vert^2}{(a_i-a_j)^2} + V(A)
	~.
}
Tracing both sides of Eq.\ \eqref{LSM:EOMs} we deduce a necessary condition on the classical solution,
\equ{
	\Tr\, V'(\Phi_\text{cl}) = 0
	~.
}
This in turn constrains the eigenvalues of $\Phi_\text{cl}$ (encoded by Eq.\ \eqref{LSM:EigenvalueMatrix}) on the real line 
to form mirror pairs around the origin. 
In detail, for
\begin{align}
\label{LSM:NecessaryConditionOnEigevalues}
\SU{2k} \,\text{ theory}&:\quad A_\text{cl} = \diag\left(a_1,-a_1,...,a_k,-a_k\right)~, \quad \text{while for}
\\[1ex]
\SU{2k+1} \,\text{ theory}&:\quad A_\text{cl} = \diag\left(a_1,-a_1,...,a_k,-a_k,0\right)
~.
\nonumber
\end{align}

\begin{table}
	\centering
	\renewcommand*{\arraystretch}{1.1}
	\begin{tabular}{|l||c|c|c|c|}
		\hline
		& \multicolumn{4}{c|}{Classes of fixed points in the \ir}
		\\
		matrix models& \textbf{Wilson-Fisher} & \textbf{Wilson-Fisher-like} & \textbf{multi-charge} & \textbf{more generic}
		\\ 
		in $2+1$ dim & $n_h=1$ & $n_h=1$ & $n_h=\floor{N/2}$ & $n_h=?$  
		\\ \hline\hline
		$\SU2$ & $\checkmark$ & $-$ & $-$ & $-$ 
		\\ \hline
		$\SU3$ & $-$ & $\checkmark$ & $-$ & $-$ 
		\\ \hline
		$\SU N$, $4\leq N\leq7$ &  $-$ & $\checkmark$ & $\checkmark$  & $-$ 
		\\ \hline
		$\SU N$, $N\geq8$ &  $-$ & $\checkmark$ & $\checkmark$  & $\checkmark$
		\\ \hline
	\end{tabular}
	\caption{The table presents the qualitatively 
		distinct classes of fixed points that appear in 
		adjoint models.
		The last column refers to  additional fixed points with possibly different behavior not classified in this work.
		For a given $N$, 
		a hyphen means that this type of fixed point cannot appear.
		$n_h$ gives the number of 
		independent $\U1$ scales admissible in the non-linear effective theory. 
	}
	\label{tb:FixedPointStructure}	
	\renewcommand*{\arraystretch}{1}
\end{table}

Eventually, using the Lax formalism (see e.g.\ \cite{Loukas:2017lof} for an application in the matrix-model setting) we find that the homogeneous and static solution to the classical \textsc{eom}s \eqref{LSM:EOMs} has two distinct branches, depending on the values of the Wilsonian parameters $g_i$. 
In both cases, there exists always a gauge where the classical solution to Eq.\ \eqref{LSM:EOMs} is parametrized as 
\equ{
	\label{LSM:ClassicalSolutionSCHEMA}
	\Phi_\text{cl} = \Ad\left[\,\exp\left(i \sum\nolimits_{j=1}^{n_h} \mu_j\, h^j\, t\right)\right]  \Phi_0
	~,
}
where $\Phi_0 \in \su N$ denotes the time-independent part. 
The direction of the time-dependence
can be conveniently taken w.l.g.\ inside the Cartan sub-algebra of $\su N$.
In this notation, $h^j\in\mathcal C_{\su N}$ are $n_h$ linearly independent directions associated to chemical potentials $\mu_j$.
As we outline in Section~\ref{ssc:SymmetryDispersions}, the corresponding embedding of the time-dependent vacuum expectation value (\vev) dictates the explicit symmetry breaking pattern in our matrix model due to non-vanishing chemical potentials. 
Modulo accidental enhancements at  special charge configurations,
the dimension $n_h$ 
gives
the number of relativistic (with linear dispersion relation) Goldstones $\chi_j$ and associated charge scales $Q_j$
in the low-energy theory. 
Consequently, $n_h$ relates the present linear description 
to the non-linear sigma models surveyed in Section~\ref{sc:NonLinear}.

\subsubsection*{The fixed-point structure of matrix theories}

In fact, the two branches of the classical solution mentioned in the previous paragraph are associated to different 
fixed points of the \textsc{rg} flow. 
Quantizing the fluctuations on top of the corresponding vacua leads to distinct predictions for the low-energy spectrum and the anomalous dimension of scalar operators.
Table~\ref{tb:FixedPointStructure} summarizes the relevant fixed-point structure for adjoint $\SU N$ theories in $2+1$ dimensions, 
based on the discussion that follows. 
It is crucial to realize that any analytic classification performed in this context is done to leading orders in the large scale in the sense of Eq.\ \eqref{Intro:SchematicEnergyExpansion}.

%
\paragraph{Multi-charge fixed point.}
Specifically, provided a $\SU{2k}$ or $\SU{2k+1}$ matrix model with $k\geq2$ there exists a fixed point  for \textit{generic} values of the couplings $g_i$ in $V(\Phi)$ (well inside the allowed parameter range). 
In Section~\ref{sc:TwoMus} we show that this class of fixed points is generally characterized by $k$  different chemical potentials in the embedding of Eq.\ \eqref{LSM:ClassicalSolutionSCHEMA}, i.e.\ $n_h=k$.
As we argue in Section~\ref{ssc:MultiChargeFixedPoint_SUN}, it generically leads to $k$ relativistic Goldstones, thus enabling us to fix up to $k=\floor{N/2}$ different $\U1$ charges in the low-energy description, while still being homogeneous in space. In the large-charge expansion up to order one, it gives a distinct prediction ($f_P\neq0$ in Eq.\ \eqref{Intro:SchematicEnergyExpansion} for $P=2$) compared to the other class of fixed points and to vector models.
We shall refer to this type of fixed points as the ``multi-charge fixed point'' of matrix theory.
To avoid confusion we stress that the term ``multi-charge'' does not refer to the actual charge assignment we consider, but to the  ability to fix multiple $\U1$ scales in the low-energy theory around a homogeneous vacuum.

\paragraph{Wilson-Fisher-like fixed point.}
To understand the other class of fixed points we take a closer look at how the scalar potential \eqref{LSM:ScalarPotential} behaves on the classical \textsc{eom}s for any $\SU N$ matrix model. By merit of condition \eqref{LSM:NecessaryConditionOnEigevalues}, $\Tr\, A_\text{cl}^3$ always vanishes identically and hence $V(A_\text{cl})$ does not depend on $g_3$.
Concentrating on the locus where $g_1=g_4=0$ and $g_2$, $g_3$ arbitrary, the scalar potential evaluated at the classical solution $\Phi_\text{cl}$ becomes
\equ{
	\label{LSM:ScalarPotential_WilsonFisher}
	V(\Phi_\text{cl}) = V(A_\text{cl}) = \frac{\mathcal R}{16}\,\Tr\,A^2_\text{cl} + g_2 \left(\Tr\,A^2_\text{cl}\right)^3	
	~.
}
Then, the full action $S[\Phi_\text{cl}]$ enjoys the  $O(N^2-1)$ symmetry.
Consequently, this branch of the solution follows the pattern of the classical ground state constructed for $O(N^2-1)$  vector models in \cite{Alvarez-Gaume:2016vff}.

In particular, the lowest-lying state of fixed charge admits only one $\U1$ charge scale given by $Q$, as there appears only one independent $\mu$ in Eq.\ \eqref{LSM:ClassicalSolutionSCHEMA}, meaning $n_h=1$. Here, only one relativistic Goldstone arises.
As we demonstrate in Section~\ref{sc:OneMu} the leading (up to order one) predictions derived at this fixed point cannot be \textit{qualitatively}\footnote{Our analytic classification does not exclude the possibility that a non-perturbative treatment results in different numerical values for the ignorance coefficients in the energy expansion \eqref{Intro:SchematicEnergyExpansion} among the models which qualitatively fall into the same class.}
distinguished from the results obtained 
at the Wilson-Fisher fixed point in the 
vector model theory (for $P=1$ it is always $f_P=0$ in Eq.\ \eqref{Intro:SchematicEnergyExpansion}).
The deviations due to $g_3\neq0$ enter only at the level of quantum fluctuations on top of the large-charge vacuum and are thus sub-leading (order $1/Q$) in the large-charge expansion.
Hence, we call this fixed point ``Wilson-Fisher-like''.

\paragraph{More fixed points in $\SU N$ adjoint models for $N\geq8$.}

To quickly see that larger symmetry groups can admit more types of fixed points, one has to  recall that the $\su N$ algebra is of rank $N-1$ and thus has $N-1$ independent Casimirs, from which we can form the $\SU N$-invariant monomials 
\equ{
	\label{CasimirInvariants}
	\Tr\, \Phi^n
	\quads{\text{for}}
	n=0,2,...,N
	~.	
}
Due to  necessary condition \eqref{LSM:NecessaryConditionOnEigevalues} we recognize from Eq.\ \eqref{CasimirInvariants} that the \textit{most general} scalar potential evaluated at the classical solution  $\Phi_\text{cl}$ of $\SU N$ matrix theory, with $N=2k$ or $N=2k+1$, can be parametrized at most by $k$ independent monomials
\equ{
	\Tr\, \Phi^{2j}_\text{cl} \quads{,}
	j=1,...,k
	~.
}
On the other hand, the particular scalar potential given 
in Eq.\ \eqref{LSM:ScalarPotential} evaluated at the classical solution, $V(\Phi_\text{cl})$, has three independent terms (recall that $\Tr\,\Phi_\text{cl}^3=0$). 
This means that it is sufficient to fully describe the large-charge behavior of  matrix theories with $k=1,2,3$.
%
Starting from $\SU8$ matrix models, we need to consider more general potentials which could also change the qualitative behavior of the theory at large charge.
Of course, there can appear more distinct classes of fixed points.
For those larger symmetry groups 
the aforementioned two types of fixed points 
appear in the locus of the space of Wilsonian parameters where the theory is described by the simpler potential Eq.\ \eqref{LSM:ScalarPotential}.

\paragraph{The special cases of $\SU2$ and $\SU3$.}

Concerning the previous classification of fixed points there are two special cases for $k=1$.
The $\SU2$ matrix model is the same as the $O(3)$ vector model and hence its analysis follows immediately from \cite{Alvarez-Gaume:2016vff}.
Despite that the $\SU3$ matrix model is not really equivalent to any vector model, it turns out that this matrix theory can only 
give qualitatively the same predictions as vector models.
In the $\su3$ algebra it is always possible to choose the basis of \eqref{CasimirInvariants} to be spanned by $\Tr\, \Phi^2$ and $\Tr\, \Phi^3$.
This equivalently means that we can set w.l.g.\  $g_1=g_4=0$ in the potential \eqref{LSM:ScalarPotential}.
Consequently, $\SU3$ falls automatically into the class of  Wilson-Fisher-like fixed points.
This is in full accordance with the explicit analysis performed in \cite{Loukas:2017lof}.
The upshot is that we can never fix more than one independent $\U1$ scale in the low-energy description of a model enjoying global $\SU3$ symmetry and be still homogeneous in space.
Starting from the $\SU4$ matrix model, we expect to see  non-trivial deviations among  the different classes of fixed points. This is why we eventually specialize on $N=4$, but we also comment on the generalization to arbitrary $N$.

\subsection{Symmetry breaking and dispersion relations}
\label{ssc:SymmetryDispersions}

In this paragraph we outline the procedure followed
to understand the symmetry breaking pattern at fixed charge and determine the Goldstone spectrum on top of the homogeneous vacuum $\braket{\Phi(t)}$ introduced above. 
To keep notation simple, we look at the situation with one chemical potential, the generalization to multiple $\mu_i$ being straight-forward.

Motivated from Eq.\ \eqref{LSM:ClassicalSolutionSCHEMA} at the level of the quantum theory we  write for the field operator
\equ{
	\label{LSM:RunningGoldstone}
	\Phi =\Ad\,[\,\e^{i \mu  t\, h}\,]\, \varPhi 
	~,
}
so that Lagrangian \eqref{LSM:FullAction} with any $\SU N$-invariant potential $V$ becomes in terms of $\varPhi \in \su N$
\equ{
	\label{LSM:ExplicitSymmetryLagrangian}
	\mathcal L = \tfrac12 \Tr \left(\partial_\mu\varPhi \partial^\mu\varPhi\right) 
	+ i \mu\, \Tr \left( [\varPhi,\dot\varPhi]\, h \right)
	- \sfrac12\mu^2\, \Tr\,[h,\varPhi]^2
	- V(\varPhi)
	~.
}
$h$ is an element in the Cartan subalgebra 
of $\mathfrak{su}(N)$. 
From the Lagrangian expressed in terms of $\varPhi$ in Eq.\ \eqref{LSM:ExplicitSymmetryLagrangian} we read off the actual symmetry, after the explicit and rank-preserving breaking due to non-zero $\mu$, generated by
\equ{
	\mathfrak h = \left\lbrace T^a \in \su N \,\vert\, \left[h,T^a\right]=0 \right\rbrace
	\quad \text{ with } \quad
	\text{rank}\,\mathfrak h  = \text{rank}\,\su N
	~.
}
In order that the \vev of the quantum operator $\braket{\Phi}$ reduces to the classical solution \eqref{LSM:ClassicalSolutionSCHEMA},
we additionally perform a spontaneous symmetry breaking by assigning the time-independent \textsc{vev} to $\varPhi$,
\equ{
	\braket{\varPhi} = \Phi_0
	~.
}
This breaks $\mathfrak h$ into $\mathfrak h'$ with
\equ{
	\mathfrak h' = \left\lbrace {T'}^{c} \in \mathfrak h \,\vert\, \left[\Phi_0, {T'}^{c}\right]=0 \right\rbrace
	~. 
}
The Goldstone counting results from the number of broken generators $\Sigma^i \in \mathfrak h$ with $[\Sigma^i,\Phi_0]\neq0$, i.e.\
\equ{
	\# \text{Goldstones} = \text{dim}\,\mathfrak h - \text{dim}\,\mathfrak h'
	~.
} 

Next, we construct the coset in order to describe the quantum fluctuations on top of $\braket{\Phi(t)}$.
Our primary objective is to find the dispersion relations for the low-energy spectrum. 
Following the standard procedure we have
\equ{
	\label{LSM:CosetConstruction}
	\Phi = \Ad [\e^{i \mu t h}]\, \varPhi = \Ad [\e^{i \mu t h}]\, \Ad[U_G]\,\Ad[U_\varphi] \left(\Phi_0 + \Phi_\text{radial}\right)
	~.
}
Note that we can rearrange the coset factors, with the corrections being of higher order in the field expansion (and hence of order $1/Q$ in the large-charge expansion). 
The Goldstone fields corresponding to true symmetries of Eq.\ \eqref{LSM:ExplicitSymmetryLagrangian} are included in 
\equ{
	\label{LSM:GoldstoneCosetFactor}
	U_G = \exp \left(i \frac{\chi_i}{v}\, \Sigma^i \right)
	\quads{\text{with}}
	[\Sigma^i,h]=0 \,\text{ and }\, [\Sigma^i,\Phi_0]\neq0 
	~,
}
while the spectator fields (which are generically expected to lead to massive modes) are included in
\equ{
	\label{LSM:SpectatorCosetFactor}
	U_\varphi = \exp \left(i \frac{\varphi_a}{v}\,N^a \right)
	\quads{\text{with}}
	[N^a, h ] \neq 0 \,\text{ and }\, [N^a, \Phi_0 ] \neq 0
	~.
} 
Here $v$ with $[v]=1/2$ denotes the characteristic field scale, $\Phi_0 \sim \order{v} \sim \order{Q^{1/4}}$ (the large-charge scaling will be justified via the explicit analysis that follows), and is used to give the proper dimensionality to the fluctuating modes in the coset.
The radial modes 
are given by 
\equ{
	\label{LSM:RadialModes}
	\Phi_\text{radial} = r_\ga\, R^\ga
	\quads{\text{with}}
	[R^\ga,\Phi_0]=0
	~.
}
Stability of the large-charge construction --\,meaning that we are expanding around a true minimum of the theory\,-- implies that any radial mode is expected to be (very) massive.

Our main task is to expand the Lagrangian \eqref{LSM:FullAction} of our linear sigma model in the fluctuations described by Eq.\ \eqref{LSM:CosetConstruction} around $\braket{\Phi(t)}$,
\equ{
	\label{LSM:LagrangianExpansion_Formal}
	\mathcal L[\Phi] = \mathcal L^{(0)}[\Phi_0\,,\,\mu] + \sum_{m\geq1} \mathcal L^{(m)}[\varPhi]
	~,
}
where $\mathcal L^{(m)}$ denotes the Lagrangian piece which is of $m$-th order in the fluctuating fields.
From the quadratic piece $\mathcal L^{(2)}$ one can read off the 
inverse propagator in momentum space, $D^{-1}(k)$ with $k\equiv k_\mu=(k_0,-\textbf k)$.
In field space it is represented by a $(N^2-1)\times(N^2-1)$ matrix 
for $\SU N$ adjoint models. 
Taking afterwards the determinant of the inverse propagator 
and  determining the roots of the resulting polynomial in $k_0$, 
\equ{
	\label{LSM:RootsOfInversePropagator}
	\det D^{-1}(k) \overset{!}{=} 0
	~,
}
gives semi-classically the desired dispersion relations in flat space.
Finally, we need to assert that the expansion \eqref{LSM:LagrangianExpansion_Formal} is well-defined, i.e.\ the higher vertices encoded in
$\mathcal L^{(m\geq3)}$ are 
controlled by $1/Q$.

\section{The multi-charge fixed point}
\label{sc:TwoMus}

The considerations in the previous section can be applied to any  theory with $\SU N$ global symmetry and matter in the adjoint representation.
From this paragraph on, we focus on the $\SU4$ matrix theory to outline a couple of novel features compared to vector models. 
The main result is summarized in Table~\ref{tb:GenericFP:Summary} and consists of the energy expansions \eqref{LSM:OneRho:AnomalousDimension}, \eqref{LSM:GenericRhos:AnomalousDimension} and \eqref{LSM:TwoRhos:AnomalousDimension} corresponding to different charge configurations depicted in Figure~\ref{fg:Q1Q2space}.
At the end of the section we also give how the $\SU N$ matrix theory should behave for any $N$ at the particular fixed point.

\subsection{Ground state and anomalous dimension at multiple charges}

In this section we discuss the first branch of the homogeneous solution to the  Euler-Lagrange \textsc{eom}s in Eq.\ \eqref{LSM:EOMs} at lowest energy and finite charge density. We are at a generic point in the space of Wilsonian parameters $\lbrace g_i\rbrace$, but well inside the cone where $V(\Phi)$ is bounded from below.

For $\SU4$ matrix models
the 
homogeneous solution to the \textsc{eom}s at the multi-charge fixed point is given by 
\equ{
	\label{LSM:TwoMus:ClassicalSolution}
	\Phi(t)  = \frac{v}{\sqrt2}
	\begin{pmatrix}
		0 & \e^{i\mu_1 t}\,\cos\frac{\vartheta}{2} & 0 & 0\\
		\e^{-i\mu_1 t}\,\cos\frac{\vartheta}{2} & 0 & 0 & 0\\
		0 & 0 & 0 & \e^{i\mu_2 t}\,\sin\frac{\vartheta}{2}\\
		0 & 0 & \e^{-i\mu_2 t}\,\sin\frac{\vartheta}{2} & 0
	\end{pmatrix}
	~,
}
up to global $\SU 4$ transformations.  The chemical potentials are fixed according to
\begin{align}
\label{LSM:ChemicalPotentials}
\mu_{1/2} \equiv\mu_{\pm} =
\sqrt{\gl_1 v^4 + \sfrac14 \left(\sfrac14\cos2\vartheta\pm\cos\vartheta\right)  \gl_2 v^4 + \sfrac{1}{8}\mathcal R }
~,
\end{align}
in terms of the effective couplings
\equ{
	\label{LSM:EffectiveCouplings}
	\gl_1 \equiv \sfrac{9}{16}g_1 + 6 g_2 + \sfrac{7}{4} g_4
	\qand
	\gl_2 \equiv 3 g_1 + 4 g_4
	~.
}
Due to the form of Eq.\ \eqref{LSM:ScalarPotential} we would generally expect three of the initial couplings to appear in the classical solution, $g_1$, $g_2$ and $g_4$.
In fact, this is the case in $\SU N$ matrix models starting from $N\geq5$.
The reason that only two couplings appear in Eq.\ \eqref{LSM:EffectiveCouplings} is special to $\SU4$ matrix theory.
The $\su4$ algebra has  only three independent Casimirs. 
In other words, the $\SU4$ invariant monomials in $V(\Phi)$ are related via \cite{MathCasimirsEtc}
\equ{
	\label{SU4CasimirsRelation}
	\Tr\,\Phi^6 + \frac18 \left(\Tr\,\Phi^2\right)^3 -\frac13 \left(\Tr\,\Phi^3\right)^2 - \frac34\, \Tr\,\Phi^4\, \Tr\, \Phi^2 = 0
	~.
}
The $\Tr\, \Phi^3$ monomial associated to $g_3$ vanishes once $\Phi=\Phi_\text{cl}$.
There thus remain two independent Casimirs, one is given by $\Tr\,\Phi^2$ and the second is found by solving Eq.\ \eqref{SU4CasimirsRelation} in terms of either $\Tr\,\Phi^6$ or $\Tr\,\Phi^4\, \Tr\, \Phi^2$\,. 
One can equivalently say that w.l.g.\ $g_4$ (or $g_1$) can be set to zero. 

In the spectral decomposition of Eq.\ \eqref{PhiEigendecomposition} the classical solution can be recast into
\equ{
	\label{LSM:TwoMus:SU4ClassicalPhi}
	\Phi(t)  = \Ad\left[
	\e^{i\left(\mu_1 h^1 + \mu_2 h^2\right)t} 
	\right]
	\Ad[U_0] A_\text{cl}
}
in terms of the 
eigenvalue matrix 
\equ{
	\label{LSM:TwoMus:EigenvalueMatrixSU4}
	A_\text{cl} = \frac{v}{\sqrt2}\, \diag \left(\cos \frac{\vartheta}{2}, - \cos \frac{\vartheta}{2} , \sin \frac{\vartheta}{2} , - \sin \frac{\vartheta}{2}\right)
	~,
}
the unitary transformation matrix
\equ{
	\label{LSM:TwoMus:TrafoMatrixSU4}
	U_0 = \frac{1}{\sqrt{2}}
	\begin{pmatrix}
		1 & -1 & 0 & 0 \\
		1 & 1 & 0 & 0 \\
		0 & 0 & 1 & -1 \\
		0 & 0 & 1 & 1 
	\end{pmatrix}
	~,
}
and the two directions in the Cartan subalgebra
\equ{
	h^1=\diag(1,-1,0,0) \qand h^2=\diag(0,0,1,-1)
	~.
}
Therefore, there are two directions ($n_h=2$ in Eq.\ \eqref{LSM:ClassicalSolutionSCHEMA})
to characterize the time-dependence of the classical background.
%
Using the particular form of $A_\text{cl}$ the original potential \eqref{LSM:ScalarPotential} becomes
\equ{
	V(\Phi(t)) = V(A_\text{cl}) = \frac{\gl_1}{6}  v^6 + \frac{\gl_2}{16} \left(\frac13 + \frac12 \cos2\vartheta \right) v^6
	~.
} 
Avoiding that $V(A_\text{cl})$ has a runway behavior when $v\rightarrow\infty$ requires  
\equ{
	\gl_1 > 0 \qand \gl_2 \in \mathbb{R}\quads{\text{with}} 
	\gl_1 \,>\, \frac{\gl_2}{16} \,>\, -\frac{\gl_1}{5}
	~.
}
Once the fluctuations on top of $\braket{\Phi(t)}$ are considered, we  find that stability of the large-charge expansion constrains this interval further to 
\equ{
	\label{TwoMus:AllowedCouplingRegion}	
	16\gl_1 \,>\, 3\gl_2 \,>\, 0
	~.
}
Within this validity region, to avoid a redundant description of the classical solution we only need to look at the first Weyl chamber, 
\equ{
	\label{LSM:TwoMus:FirstWeylChamber}
	\vartheta \in [0,\pi/2] 
	~.
}

\begin{figure}
	\centering
	\includegraphics[width=9cm,height=9cm,keepaspectratio]{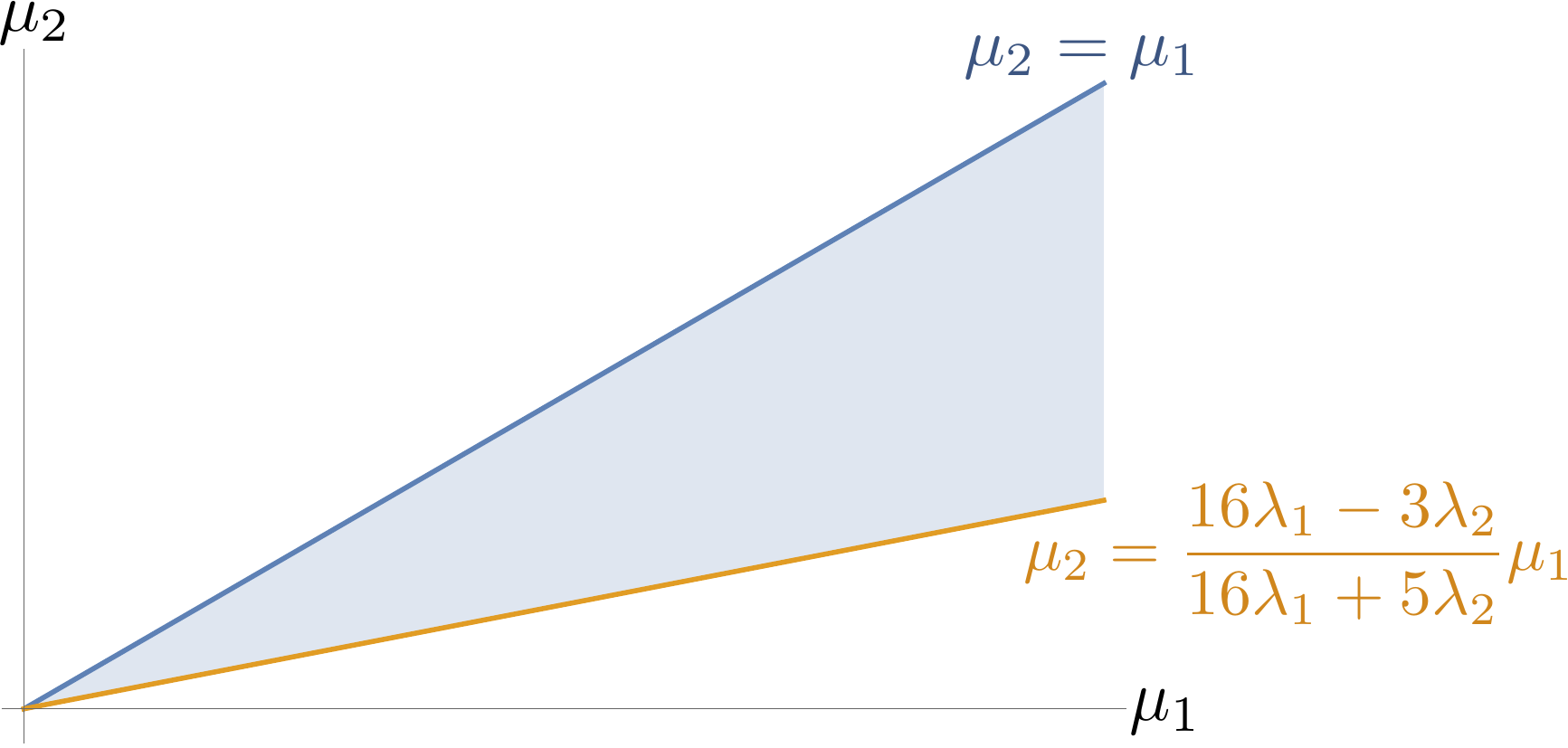}
	\caption{The shaded region indicates the possible values of the chemical potentials $\mu_1$ and $\mu_2$ parametrizing the time-dependence of the classical ground state, Eq.\ \eqref{LSM:TwoMus:SU4ClassicalPhi}, in the first Weyl chamber.}\label{fig:MusExclusion}
\end{figure}

Generally, one could expect $\mu_1$ and $\mu_2$ to be arbitrary. However, for a simple Lie group there are  constraints imposed by Eq.\ \eqref{LSM:ChemicalPotentials}, meaning that
not all points in the space spanned by $h^1$ and $h^2$ can be reached, but only those satisfying
\equ{
	\label{LSM:TwoMus:ChemicalPotentialsRelation}
	\mu_2 = \mu_1\, \sqrt{1-\frac{8 \lambda _2 \cos \vartheta}{16 \lambda_1+4 \lambda_2 \cos\vartheta+\lambda_2\cos 2 \vartheta }}
	\,+\,
	\text{subleading $\mathcal R$-dependent terms}
	~.
}
The admissible $(\mu_1,\mu_2)$ tuples within the first Weyl chamber  \eqref{LSM:TwoMus:FirstWeylChamber} (where $\mu_1\geq\mu_2\geq0$) are plotted in Figure~\ref{fig:MusExclusion}.
The chemical potentials used here to parametrize the time-dependence of the general solution \eqref{LSM:TwoMus:ClassicalSolution} should not be confused with the chemical potentials we are going to use in Section~\ref{sc:NonLinear} to write the non-linear sigma models.
The former see the full structure of the global symmetry algebra $\su4$, whereas the latter know only about the low-energy dynamics described by the relevant coset spaces which we derive in Sections~\ref{ssc:TwoMus_OneU1} and \ref{ssc:TwoMus_TwoRhos}. 
%

%
\paragraph{The Noether current.} In the chosen gauge the Noether matrix \eqref{LSM:NoetherMatrix} is conveniently diagonal,
\begin{align}
\label{LSM:NoetherCurrent:TwoMus}
J_0 =
i  [\Phi,\dot \Phi]
= v^2\,\diag \left(\mu_1 \cos^2\frac\vartheta2\,,\,-\mu_1 \cos^2\frac\vartheta2\,,\,\mu_2  \sin^2\frac\vartheta2\,,\,-\mu_2 \sin^2\frac\vartheta2\right)
~.
\end{align}
Charge conservation  applied to our homogeneous state, $\dot J_0=0$,  implies that both $v$ and $\vartheta$ are constant.
Furthermore,
if at least one global charge is taken sufficiently large it becomes  clear that the radial condensate $v$ is also large.
It is important to stress that for generic $\vartheta$ and in the allowed region \eqref{TwoMus:AllowedCouplingRegion} it is $\mu_1\neq\mu_2$ as manifested by relation \eqref{LSM:TwoMus:ChemicalPotentialsRelation}. 
In the low-energy theory, this will allow us to independently fix  
two  $\U1$ scales, $Q_1$ and $Q_2$, characterized 
by charge densities $\rho_1=Q_1/\mathcal V$ and $\rho_2=Q_2/\mathcal V$ respectively. 
Thus, we take in our linear description
\equ{
	\label{LSM:TwoMus:NoetherCondition}
	B\, J_0 \overset{!}{=} \diag \left(\rho_1\,,\,-\rho_1\,,\,\rho_2\,,\,-\rho_2\right)
	\quads{\text{with}}
	B = \diag\left(b_1,b_1,b_2,b_2\right)
	~.
}
$b_1$ and $b_2$ are order-one parameters chosen such that the associated conserved global charges
\equ{
	\label{Def:GlobalCharge}
	Q_i = \int_{\Sigma} \d \Sigma\, \rho_i 
	\,\in \, \mathbb Z
	\quads{,}
	i=1,2~,
}
are properly quantized, independently of the global properties of the field $\Phi$.
They depend on the microscopic structure of a given physical system.
To better comprehend how they arise we need to investigate the contribution of higher terms to the linear Lagrangian \eqref{LSM:FullAction}.
Due to Lorentz- and $\SU N$-invariance
they  always take the form
\equ{
	\frac{\Tr \left(\partial_\mu\Phi\partial^\mu\Phi\cdots \Phi\cdots \partial_\nu\Phi\partial^\nu\Phi \cdots\right)}{(\Tr\,\Phi^2)^\#}
	~,
}
where the denominator is chosen such that the higher operator has mass dimension three in three space-time dimensions.
On the classical solution $\Phi=\Phi_\text{cl}$ specified by Eq.\ \eqref{LSM:TwoMus:ClassicalSolution} the contribution
of all these higher vertices changes 
the angular momentum matrix $K$ defined in Eq.\ \eqref{AngularMomentumInTheBody} to $BK$. This in turn implies together with $[B,U]=0$ for the transformation matrix of Eq.\ \eqref{LSM:TwoMus:TrafoMatrixSU4} that the Noether matrix $J_0$ is modified to $BJ_0$ as indicated in \eqref{LSM:TwoMus:NoetherCondition} with
\equ{
	\label{LSM:TwoMus:Bmatrix}
	b_1 = \sum_{n,m=0}^\infty c_{nm}\left(\frac{\mu_1}{v^{2}}\right)^n \cos^m \frac{\vartheta}{2}
	\qand
	b_2 = \sum_{n,m=0}^\infty c_{nm}\left(\frac{\mu_2}{v^{2}}\right)^n \sin^m \frac{\vartheta}{2}
	~,
}
where $c_{nm}$ are infinitely many  Wilsonian parameters of order one.
Thus, $b_1$ and $b_2$ are in fact functions of $\vartheta$.
In the first Weyl chamber it follows that $b_1\geq b_2\geq0$.

Combining Eq.\ \eqref{LSM:NoetherCurrent:TwoMus} and \eqref{LSM:TwoMus:NoetherCondition} into
\equ{
	\label{LSM:SU4ChargesRelation}
	\rho_2  = \rho_1\, \frac{b_2}{b_1}\tan^2\frac{\vartheta}{2}\, \sqrt{1-\frac{8 \lambda _2 \cos \vartheta}{16 \lambda_1+4 \lambda_2 \cos\vartheta+\lambda_2\cos 2 \vartheta }}
	\,+\, \order{\frac{\mathcal R}{\rho_1}}
	~,
}
we recognize that $Q_1 \gg1$ can be used w.l.g.\ as a good expansion parameter in this setting.
In the first Weyl chamber \eqref{LSM:TwoMus:FirstWeylChamber}, where the $\vartheta$-dependent factor takes values from 0 to 1, it is $\rho_1 \geq \rho_2\geq0$.
Eq.\ \eqref{LSM:SU4ChargesRelation} is very important because it relates the angle $\vartheta$ to the ratio of the fixed charges. 
Due to the  $\vartheta$-dependence of $b_1$ and $b_2$ it is not possible however to analytically solve the relation for  $\vartheta$.
As we are going to see in the following, this inability will result in 
infinitely many ignorance coefficients in the large-charge expansion, two for each possible value of the ratio $Q_2/Q_1$.

\paragraph{The condensate energy.}
Independent of the charges $Q_1$ and $Q_2$ fixed in Eq. \eqref{LSM:TwoMus:NoetherCondition} the energy of the classical ground state can be always given as a perturbative expansion in some suitably chosen large parameter.
As we commented below \eqref{LSM:SU4ChargesRelation} this is naturally chosen to be $Q_1 \gg1$.
Then  
using that $0\leq \cos\vartheta \leq 1$, 
we easily deduce the relevant scalings 
\equ{
	v\sim\order{Q^{1/4}_1} \qand
	\mu_1,\mu_2 \sim \order{Q^{1/2}_1}
	~.
}
Implementing those scalings, we generically find from the Hamiltonian \eqref{LSM:HomogeneousHamiltonian} the energy of the condensate on a compact manifold with volume $\mathcal V$, 
\begin{align}
\label{LSM:TwoMus:CondensateEnergy}
E_0 = &\,\,
\frac{1}{2}\mu_1^2  v^2 \cos^2 \frac{\vartheta}{2}
+\frac{1}{2}\mu_2^2  v^2 \sin^2 \frac{\vartheta}{2}
+\frac{\mathcal R}{16} v^2
+
v^6 \left(\frac{\lambda 1}{6}+\frac{\lambda_2}{48}+\frac{\lambda_2}{32}  \cos 2 \vartheta\right)
\\[0.8ex]
=&\,\,
\left(\frac{1}{b_1(\vartheta)}\right)^{3/2}
\frac{16 \lambda_1+2\lambda_2 + 3 \lambda _2 \cos 2 \vartheta}{3\cos^3(\vartheta/2) \left(16 \lambda_1 + 4 \lambda_2\cos\vartheta +\lambda_2 \cos 2 \vartheta\right)^{3/4}}
\left(\frac{Q_1}{\mathcal V}\right)^{3/2}
\nonumber
\\[1ex]
+&\,\,
\left(\frac{1}{b_1(\vartheta)}\right)^{1/2}
\frac{16 \lambda_1-2 \lambda_2+8 \lambda_2 \cos\vartheta -\lambda_2 \cos 2\vartheta}{8 \cos(\vartheta/2)\left(16 \lambda_1+4 \lambda_2 \cos\vartheta +\lambda_2 \cos2 \vartheta\right)^{5/4}} \,\mathcal R
\left(\frac{Q_1}{\mathcal V}\right)^{1/2}
+
\order{Q_1^{-1/2}}
~,
\nonumber
\end{align}
as an asymptotic expansion in $1/Q_1 \ll 1$. 
In this formula $\vartheta$ is determined by the exact solution of Eq.~\eqref{LSM:SU4ChargesRelation}.
Hence,
the order-one coefficients in front of the $Q^{3/2}_1$- and $Q^{1/2}_1$-term 
depend 
on the underlying model as described 
by the Wilsonian parameters and on the ratio of the two charges $Q_2/Q_1$. 

\begin{figure}[t]
	\centering
	\includegraphics[width=11cm,height=11cm,keepaspectratio]{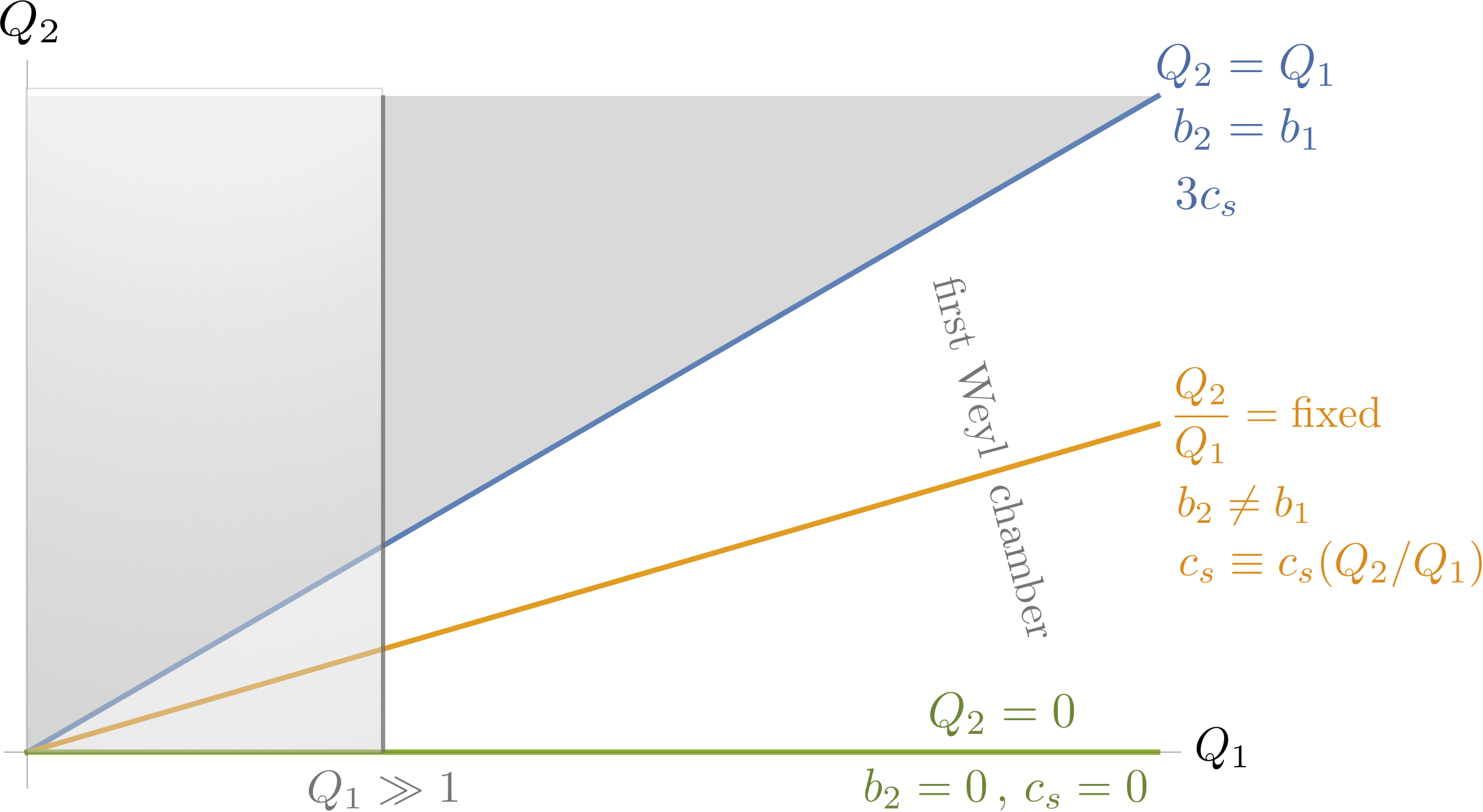}
	\caption{The relation \eqref{LSM:SU4ChargesRelation} is plotted in the first Weyl chamber where  $Q_1\geq Q_2\geq0$. To each fixed ratio $Q_2/Q_1$ with $Q_1\gg1$ there generically corresponds an orange line along which the expansion \eqref{LSM:AnomalousDimension} has the same ignorance coefficients and the same non-universal contribution $c_s$ at order one (see also Table~\ref{tb:GenericFP:Summary}). 
		The blue and green lines  describe limiting configurations of enhanced symmetry, analyzed in Section~\ref{ssc:TwoMus_OneU1} and at the end of \ref{ssc:TwoMus_TwoRhos}.}\label{fg:Q1Q2space}
\end{figure}

\paragraph{Anomalous dimension.} By the standard state-operator correspondence in any \textsc{cft} we can map the vacuum energy $\braket{E}=E_0$ for the constructed state at fixed charges $Q_1$ and $Q_2$ on the manifold $\Sigma$ to the anomalous dimension $\Delta$ of the lowest scalar operator with the same charge configuration 
in $\mathbb R^3$  space.
Concretely, we only need to take the compact manifold to be the unit two-sphere, $\Sigma=S^2$, which means substituting $\mathcal V = 4\pi$ and $\mathcal R = 2$ in Eq.\ \eqref{LSM:TwoMus:CondensateEnergy}. Then, we automatically obtain the 
condensate contribution $E_0\vert_{S^2}$ to the anomalous dimension $\Delta(Q_1,Q_2)$.

Including the order-one contribution $E_\text{Casimir}^G(S^2)$ %
due to the Casimir energy on the two-sphere of relativistic Goldstones that follows after analyzing the quantum fluctuations   
in the subsequent section, the leading prediction\footnote{As long as these leading terms are concerned, the action~\eqref{LSM:FullAction} is sufficient to capture all the physics. Including any higher-derivative terms in this action will simply change the ignorance coefficients $c_{3/2}$ and $c_{1/2}$ but will not spoil the form of the $Q_1$-expansion and most importantly the prediction at order one.} for the anomalous dimension is given by
\equ{
	\label{LSM:AnomalousDimension}
	\begin{align}
	\Delta(Q_1,Q_2)=&\,\, E_0\vert_{S^2} 
	+ E_G^\text{Casimir}(S^2) + \order{Q_1^{-1/2}}
	\\
	=&\,\, c_{3/2}(Q_2/Q_1) \left(\frac{Q_1}{4\pi}\right)^{3/2} + c_{1/2}(Q_2/Q_1) \left(\frac{Q_1}{4\pi}\right)^{1/2}
	- 0.0937 - f_2(Q_2/Q_1) + \order{Q_1^{-1/2}}
	,
	\nonumber
	\end{align} 
}
as an expansion in $1/Q_1\ll1$.
$f_2(Q_2/Q_1)$ summarizes the contribution from relativistic Goldstones with non-universal speeds of sound $c_s$ given in Eq.\ \eqref{LSM:OneRho:SecondGoldstone_SpeedOfSound} and \eqref{LSM:GenericCase:SpeedsOfSound}.
In Figure~\ref{fg:Q1Q2space} we indicate how this formula should behave along various charge configurations in the first Weyl chamber where $Q_1\geq Q_2$.

Specifically, for each fixed ratio $Q_2/Q_1$ we obtain an asymptotic expansion in $1/Q_1$ with (the same for all $Q_1$) ignorance coefficients $c_{3/2}$ and $c_{1/2}$.
%
Therefore, along each orange line in Figure~\ref{fg:Q1Q2space} there exists a meaningful large-charge expansion  whose leading orders are dictated by the condensate energy \eqref{LSM:TwoMus:CondensateEnergy} on the two-sphere.
Notice that this expansion is qualitatively the same as the one encountered at a  fixed point of Wilson-Fisher type, cf.\ formula \eqref{LSM:WislonFisher:AnomalousDimension}.
Expanding $\Delta$ along some different line (i.e.\ for a different fixed ratio $Q_2/Q_1$) leads to a distinct large-charge prediction, already to leading orders,
in the sense that the ignorance coefficients $c_{3/2}$ and $c_{1/2}$ are generically expected to be different.
This is due to the infinite series of corrections summarized by $b_1$ and $b_2$ in Eq.\ \eqref{LSM:TwoMus:Bmatrix}.
%

In addition to the non-trivial situation encountered for the condensate part $E_0\vert_{S^2}$, formula \eqref{LSM:AnomalousDimension} also gives different predictions at order one deepening on the particular charge assignment and the associated symmetry-breaking pattern.
The study of such predictions is the subject of the semi-classical as well as quantum analysis that follows.

\subsection{A charge configuration with enhanced symmetry}
\label{ssc:TwoMus_OneU1}

It is instructive to first discuss the limiting case $Q_1=Q_2$ of the fixed-charge configuration to better comprehend the concepts outlined in Section~\ref{ssc:SymmetryDispersions} as well as to clearly demonstrate the novel features arising at the present fixed point.
Ultimately, we analyze  the general case with $Q_1 \neq Q_2$.
Our objective is to read off the Goldstone spectrum on top of the vacuum state 
$\braket{\Phi(t)}$ and compute the associated Casimir energy $E_G^\text{Casimir}(S^2)$ on the two-sphere. 
This enables us to provide a meaningful perturbative expansion (Eq.\ \eqref{LSM:GenericRhos:AnomalousDimension} together with its limiting scenarios Eq.\ \eqref{LSM:OneRho:AnomalousDimension} and  \eqref{LSM:TwoRhos:AnomalousDimension}) for the anomalous dimension 
of the lowest scalar operator. 
The overall outcome of the semi-classical analysis that follows is summarized in Table~\ref{tb:GenericFP:Summary}.

We consider the extreme case with  $Q_1=Q_2$ to perform perturbative expansions in $Q_1\gg1$.
This saturates the upper limit within the first Weyl chamber (blue line in Figure~\ref{fg:Q1Q2space}).
Contrary perhaps to naive expectation, fixing one charge scale still has tractable consequences in the fluctuations on top of $\braket{\Phi(t)}$ in this branch of the classical solution. 
In such a charge configuration $\vartheta=\pi/2$, so that 
the unique large-charge scale $Q_1 = 4\pi\rho_1$ is associated to the Noether current matrix via Eq.\ \eqref{LSM:TwoMus:NoetherCondition}:
\equ{
	J_0 
	\overset{!}{=} \frac{\rho_1}{b_1} \diag \left(1,-1,1,-1\right)
	\quads{\text{with}}
	\frac{\mu v^2}{2} = \frac{\rho_1}{b_1}
	~,
} 
where
the chemical potentials in the classical solution \eqref{LSM:TwoMus:ClassicalSolution} are naturally identified, 
\equ{
	\label{ChemicalPotentials_Identified}
	\mu \equiv \mu_1 = \mu_2 
	=
	\frac{1}{\sqrt{2}}(16\lambda_1-\lambda_2)^{1/4}\sqrt{\frac{\rho_1}{b_1} }
	\,+\, \order{\rho_1^{-1/2}}
	~,
}
and from Eq.\ \eqref{LSM:TwoMus:Bmatrix} it also follows that $b_1=b_2$. 
The radial amplitude in the time-independent \textsc{vev} $\Phi_0$ scales with
\equ{
	v 
	=
	\left(\frac{64}{16 \lambda_1-\lambda_2}\right)^{1/8} \left(\frac{\rho_1}{b_1}\right)^{1/4}
	+\, 
	\order{\rho_1^{-3/4}}
	~.
}
Evidently, both $v$ and $\mu$ are large, when $\rho_1$ is large, so they can be used as expansion parameters, as well.
This is a mere technical convenience; eventually, everything is expressed in terms of the one large scale, the global charge $Q_1=Q_2$.

The symmetry breaking pattern in this extreme case reads 
\equ{
	\label{LSM:ToyModel:SymmetryBreaking}
	\SU4 ~\xrightarrow{\text{explicit}}~  \U2 \times \SU2~\xrightarrow{\text{spontaneous}}~ \SU2'
	~.
}
In detail, the Cartan generator of the time-dependent \textsc{vev} in Eq.\ \eqref{LSM:RunningGoldstone},
\equ{
	h = \diag \left(1,-1,1,-1\right)
	~,
}
leaves unbroken the two $\su2$ subalgebras generated by
\renewcommand*{\arraystretch}{0.9}
\begin{align}
T^1=&\,
\begin{pmatrix}
0&0&1&0\\
0&0&0&0\\
1&0&0&0\\
0&0&0&0
\end{pmatrix}
~,~~
T^2=
\begin{pmatrix}
0&0&i&0\\
0&0&0&0\\
-i&0&0&0\\
0&0&0&0
\end{pmatrix}
~,~~
T^3=
\begin{pmatrix}
1&0&0&0\\
0&0&0&0\\
0&0&-1&0\\
0&0&0&0
\end{pmatrix}
\\
\text{and}\quad
T^4=&\,
\begin{pmatrix}
0&0&0&0\\
0&0&0&1\\
0&0&0&0\\
0&1&0&0
\end{pmatrix}
~,~~
T^5=
\begin{pmatrix}
0&0&0&0\\
0&0&0&i\\
0&0&0&0\\
0&-i&0&0
\end{pmatrix}
~,~~
T^6=
\begin{pmatrix}
0&0&0&0\\
0&1&0&0\\
0&0&0&0\\
0&0&0&-1
\end{pmatrix}
~,
\nonumber
\end{align}
as well as the $\uu 1$ generator described by $h$ itself.
Out of these generators, the $\su2'$ subalgebra generated by
\equ{
	\label{LSM:OneRho:SurvivingSU2Prime}
	{T'}^1 = T^1+T^4
	\quads{,}
	{T'}^2 = T^2+T^5
	\qand
	{T'}^3 = T^3+T^6
}
remains unbroken by the time-independent \textsc{vev} $\Phi_0$ (obtained from \eqref{LSM:TwoMus:ClassicalSolution} for $t=0$ and $\vartheta=\pi/2$).
Hence, we are expecting the Goldstone spectrum to live in the  coset space 
\equ{
	\label{LSM:OneRho:CosetSpace}
	\left(\U2 \times \SU2 \right)/\,\SU2' = \U2
}
of dimension 4. The corresponding coset factor $U_G$ in Eq.\ \eqref{LSM:GoldstoneCosetFactor} can be subsequently parametrized by
\begin{align}
\Sigma^1 = \diag\left(1,-1,0,0\right)  
\quads{,}
\Sigma^2 = \diag \left(0,0,1,-1\right) 
\qand
\Sigma^3 = T^1
\quads{,}
\Sigma^4 = T^2
~.
\end{align}
Note that this choice (which is dictated by mere convenience in the subsequent expansion of the Lagrangian) for the generators $\Sigma^i$ is not unique, but up to identifications in the coset.
These identifications are described by  elements in the surviving $\su2^\prime$ algebra \eqref{LSM:OneRho:SurvivingSU2Prime}.
In particular, to make the underlying $\U2$ group structure of the coset space apparent, 
one starts from the provided parametrization in terms of $\Sigma^i$ and defines new generators ($\cong$ means equality in the coset space)
\begin{align}
\tilde\Sigma^1=\Sigma^3
\quads{,} &\,
\tilde\Sigma^2=\Sigma^4
\quads{,}
\tilde\Sigma^3= \diag\left(1,0,-1,0\right)=\sfrac12\left(\Sigma^1-\Sigma^2+\diag\left(1,1,-1,-1\right)\right) \cong  \sfrac12\left(\Sigma^1-\Sigma^2\right)
\nonumber
\\[1ex]
\qand &
T_{\U1} = \Sigma^1+\Sigma^2=\diag\left(1,-1,1,-1\right)
~,
\end{align}
which satisfy the $\su2$ algebra commutation relations, $[\tilde\Sigma^i,\tilde\Sigma^j]=2i\ge_{ijk}\tilde\Sigma^k$ as well as $[\tilde\Sigma^i, T_{\U1}]=0$.
On the other hand, the base part specified by Eq.\ \eqref{LSM:RadialModes} includes seven radial modes,
\begin{align}
\Phi_\text{radial} = 
\begin{pmatrix}
r_5 & r_1 & r_6+i\,r_7 & r_3+i\,r_4\\
r_1 & r_5 & r_3+i\,r_4 & r_6+i\,r_7\\
r_6-i\,r_7 & r_3-i\,r_4 & -r_5 & r_2\\
r_3-i\,r_4 & r_6-i\,r_7 & r_2 & -r_5
\end{pmatrix}
~.
\label{LSM:OneRho:RadialModesGenerators}
\end{align}
The spectator fields $\varphi_a$ parameterizing the coset factor $U_\varphi$ in Eq.\ \eqref{LSM:SpectatorCosetFactor} are aligned along
\renewcommand*{\arraystretch}{0.9}
\begin{align}
U_\varphi = \exp \frac{i}{v}
\begin{pmatrix}
0&i \varphi_1&0&0\\
-i\varphi_1&0&\varphi_3+i\varphi_4&0\\
0&\varphi_3-i\varphi_4&0&i\varphi_2\\
0&0&-i\varphi_2&0
\end{pmatrix}
~.
\end{align} 

Now, we are in a position to 
expand the Lagrangian $\mathcal L = \mathcal K - \mathcal V$ up to quadratic order in the fluctuating fields and up to order one in the chemical potential $\mu \sim \order{\sqrt{Q_1}}$: 
\begin{align}
\label{LSM:OneRho:QuadraticLagrangian}
\mathcal L =&\,\, 
\frac{16 \mu ^3-3 \mu 
	\mathcal{R}}{12 \sqrt{16 \lambda _1-\lambda _2}}
+
\frac{\sqrt{2} \mu ^{3/2} \dot{\chi }_1}{\left(16 \lambda_1-\lambda_2\right)^{1/4}}+\frac{\sqrt{2} \mu ^{3/2} \dot{\chi }_2}{\left(16 \lambda_1-\lambda_2\right)^{1/4}}
\\[0.8ex]
&\,\,
-\frac{\left(16 \lambda_1+\lambda_2\right)\left(8 \mu^2-\mathcal{R}\right)}{128\lambda_1-8 \lambda _2}r_1^2
-\frac{\left(16 \lambda_1-3 \lambda_2\right) \left(8 \mu^2-\mathcal{R}\right)}{64 \lambda_1-4\lambda_2}r_2 r_1
-\frac{\left(16 \lambda_1+\lambda_2\right) \left(8 \mu^2-\mathcal{R}\right)}{128 \lambda_1-8 \lambda_2} r_2^2
\nonumber
\\[0.8ex]
&\,\,
-\frac{\lambda_2 \left(8 \mu^2-\mathcal{R}\right)}{32\lambda_1-2 \lambda_2}r_3^2 - \frac{\lambda_2 \left(8 \mu^2-\mathcal{R}\right)}{32 \lambda _1-2 \lambda _2}r_4^2
-\frac{\left(16 \lambda _1+7 \lambda _2\right) \mu^2-\lambda_2 \mathcal{R}}{32 \lambda _1-2 \lambda _2}r_5^2
\nonumber
\\[0.8ex]
&\,\,
-\frac{\left(16 \lambda _1+7 \lambda _2\right) \mu^2-\lambda _2 \mathcal{R}}{32 \lambda_1-2 \lambda _2}r_6^2
-\frac{\left(16 \lambda_1+7 \lambda_2\right) \mu^2 -\lambda_2 \mathcal{R}}{32 \lambda _1-2 \lambda_2}r_7^2 
\nonumber
\\[0.8ex]
&\,\,
-\frac{1}{2} \mu ^2 \varphi _1^2-\frac{1}{2} \mu ^2 \varphi_2^2-\frac{1}{2} \mu ^2 \varphi_3^2-\frac{1}{2} \mu ^2
\varphi_4^2-\mu \, \varphi_3 \dot{\varphi}_4-\mu  \, \dot{\varphi}_3 \varphi_4
\nonumber
\\[0.8ex]
&\,\,
+2 \mu  r_1 \dot{\chi}_1+2 \mu  r_2 \dot{\chi}_2+2 \mu 
r_3 \dot{\chi}_3
+2 \mu  r_4 \dot{\chi}_4
\nonumber
\\[0.8ex]
&\,\,
+\sfrac12\sum_{\ga=1}^7 \left(\dot r_\ga^2 - (\nabla r_\ga)^2\right)
+
\sfrac12\sum_{a=1}^4 \left(\dot \varphi_a^2 - (\nabla \varphi_a)^2\right)
+
\sfrac12\sum_{i=1}^4\left(\dot{\chi_i}^2 - (\nabla\chi_i)^2
\right)
+\order{\mu^{-1/2}}
~.
\nonumber
\end{align}
Our coset parametrization ensures that the kinetic term is diagonal in the fluctuations $r_\ga\,,\,\varphi_a$ and $\chi_i$\,.
Furthermore, several fields have been appropriately rescaled by numerical factors such that the  normalization of the kinetic terms is canonically set to $1/2$. 
The inverse propagator $D^{-1}(k)$ in this coset parametrization takes block-diagonal form by ordering the fields as $\lbrace r_1,r_2,\chi_1,\chi_2~,~\chi_3,r_3,\chi_4,r_4,$ $r_5,r_6,r_7~,~\varphi_1,\varphi_2,\varphi_3,\varphi_4 \rbrace$: 
\begin{equation}
\label{ToyModel:PropagatorBlocks}
D^{-1}(k) = \diag\left(D^{-1}(k)\vert_{r_1,r_2,\chi_1,\chi_2}\,,\,D^{-1}(k)\vert_{\chi_3,r_3,\chi_4,r_4}\,,\,  D^{-1}(k)\vert_{r_5,r_6,r_7} \,,\, D^{-1}(k)\vert_{\varphi_1,\varphi_2,\varphi_3,\varphi_4} \right)
~.
\end{equation}
The explicit expressions for the blocks are provided in the first paragraph of Appendix~\ref{app:Propagators}.

\begin{table}[t!]
	\centering
	\renewcommand*{\arraystretch}{1.1}
	\begin{tabular}{|c|c|c|c|}
		\hline
		\textbf{propagator block} & \textbf{dispersion relation} $\omega+ \order{\mu^{-1}}$ & \textbf{multiplicity} & \textbf{type}
		\\\hline\hline
		\multirow{4}{*}{$\left.D^{-1}(k)  \right\vert_{r_1,r_2,\chi_1,\chi_2}$} & $\abs{\textbf k}/\sqrt 2$ & 1 & relativistic, universal 
		\\\cline{2-4} \rule{0pt}{3.2ex}
		& $\sqrt{\frac{2\gl_2}{16\gl_1+\gl_2} } \,\abs{\textbf k}$ & 1 & relativistic, model-dep.
		\\[0.8ex]\cline{2-4}
		& $2 \sqrt2 \,\mu $ & 1 & massive
		\\\cline{2-4}\rule{0pt}{3.2ex}
		& $2\sqrt{\frac{16 \lambda_1+\lambda_2}{16 \lambda_1-\lambda_2}}\, \mu$ & 1  & massive
		\\[0.8ex]\hline\rule{0pt}{3.2ex}
		\multirow{2}{*}{$D^{-1}(k)\vert_{\chi_3,r_3,\chi_4,r_4}$} & $\sqrt{\frac{2\gl_2}{16\gl_1+\gl_2} } \,\abs{\textbf k}$ & 2 &  relativistic, model-dep.
		\\[0.8ex]\cline{2-4}\rule{0pt}{3.2ex}
		& $2 \sqrt{\frac{16 \lambda _1+\lambda _2}{16 \lambda _1-\lambda _2}}\, \mu$ &2  & massive
		\\[0.8ex]\hline\rule{0pt}{3.2ex}
		\multirow{1}{*}{ $D^{-1}(k)\vert_{r_5,r_6,r_7}$} & $\sqrt{\frac{16\gl_1+7\gl_2}{16\gl_1-\gl_2}} \mu$ & 3 & massive
		\\[0.8ex]\hline
		\multirow{1}{*}{ $D^{-1}(k)\vert_{\varphi_1,\varphi_2,\varphi_3,\varphi_4}$} & $\mu$ & 4 & massive
		\\\hline
	\end{tabular}
	\caption{The table lists the spectrum found semi-classically in the limiting situation with $Q_1=Q_2$ and $b_1=b_2$. The propagator blocks refer to Eq.\ \eqref{ToyModel:PropagatorBlocks}. The leading dispersion relation of each mode is specified and how many times it is obtained in a given block. $\mu\sim\order{\sqrt{Q_1}}$ sets the mass scale.
	}
	\label{tb:ToyModel:Spectrum}	
	\renewcommand*{\arraystretch}{1}
\end{table}

Using these blocks to determine the roots of Eq.\ \eqref{LSM:RootsOfInversePropagator} we obtain the various gapless and massive modes listed in Table~\ref{tb:ToyModel:Spectrum}.
In detail, the light spectrum comprises four relativistic Goldstone fields.
The first relativistic Goldstone has a 
universal, model-independent dispersion relation, while the other three Goldstone modes have a model-dependent speed of sound:
\equ{
	\label{LSM:OneRho:SecondGoldstone_SpeedOfSound}
	c_{s}^{(1)}=1/\sqrt2 
	\qand
	\underbrace{c_{s}^{(2)} = \sqrt{\frac{2\gl_2}{16\gl_1+\gl_2} } < 1
		~~\text{ for }\,\, 16\gl_1>3\gl_2>0}_{3 \text{ times}}
	~.
}
Inside the allowed region \eqref{TwoMus:AllowedCouplingRegion} 
the model-dependent speed of sound 
satisfies causality.
In contrast to the $\SU3$ matrix model and the $O(n)$ vector models exhibiting only one relativistic mode with the universal speed of sound, as long as vacuum configurations are considered which are homogeneous in space, additional relativistic \textsc{dof}s emerge in our homogeneous $\SU4$ setting. 
Their dispersion relation in this setup is the same,
but non-universal, since it depends on the effective couplings $\gl_1$ and $\gl_2$, which encode microscopic information about the underlying physical model.
%
All other modes are heavy with masses which scale with $\mu\sim\order{\sqrt{Q_1}}$.
%

With the derived spectrum at hand, it is easy to see that the energy formula in Eq.\ \eqref{LSM:AnomalousDimension} receives  a tractable contribution from the fluctuations at order one.
This contribution comes from the four relativistic Goldstones $\chi_i$ with dispersion relation on the unit-sphere $S^2$ given by 
\equ{
	\omega^{(j)}(S^2) = c_s^{(j)} \sqrt{l(l+1)} + \order{1/Q_1} 
	\quads{,}
	l \in \mathbb Z
	~,
}
where the speed of sound $c_s^{(j)}$, $j=1,2$\,, is read off from 
Eq.\ \eqref{LSM:OneRho:SecondGoldstone_SpeedOfSound}.
These relativistic modes contribute to the vacuum energy 
via their Casimir energy \cite{Monin:2016bwf}:
\equ{
	\label{CasimirEnergyGoldstone_S2}
	E_\text{Casimir}^{(j)}(S^2) = \frac{c^{(j)}_s}{2} \left(-\frac14 - 0.015096\right)
	~.
}
Hence, by the state-operator correspondence ($\mathcal R=2$ on $S^2$) the final formula for the anomalous dimension of the lowest scalar operator with charges $Q_1 = Q_2$ 
becomes
\begin{align}
\label{LSM:OneRho:AnomalousDimension}
\Delta(Q_1,Q_1) =&\,\, 
\frac{2\sqrt{2}}{3}  \left(16 \lambda_1-\lambda_2\right)^{1/4} \left(\frac{Q_1}{4\pi b_1}\right)^{3/2}
+ 
\frac{1}{2\sqrt2\left(16 \lambda_1-\lambda_2\right)^{1/4}}
\left(\frac{Q_1}{4\pi b_1}\right)^{1/2}
\nonumber
\\[1ex]
&~~-0.0937
- \,{3 \times}\, 0.1325
\sqrt{\frac{2\gl_2}{16\gl_1+\gl_2}}
\,+\order{Q_1^{-1/2} }
~.
\end{align}
As anticipated, this perturbation series is of the form schematically given in Eq.\ \eqref{Intro:SchematicEnergyExpansion} with the model-dependent contribution 
given by $f_P = \,{3 \times}\, 0.1325\, c_s^{(2)}$.
Any effect due to higher vertices from the Lagrangian expansion \eqref{LSM:LagrangianExpansion_Formal} is suppressed, as shown in Section~\ref{ssc:LoopSuppression} by powers of $1/Q_1$.

\subsection{The dispersion relations for generic charges}
\label{ssc:TwoMus_TwoRhos}

In this section we deduce the spectrum when two different charges $Q_1$ and $Q_2$ are 
fixed in Eq.\ \eqref{LSM:TwoMus:NoetherCondition}.
We parametrize the fluctuations on top of the classical vacuum $\braket{\Phi(t)}$ according to Eq.\ \eqref{LSM:CosetConstruction}.
For \textit{generic} $\vartheta$ in the classical solution \eqref{LSM:TwoMus:ClassicalSolution} 
the radial modes satisfying $[\Phi_0,\Phi_\text{radial}]=0$ are given by
\equ{
	\Phi_\text{radial} = \frac{1}{\sqrt2} 
	\renewcommand*{\arraystretch}{1.1}
	\begin{pmatrix}
		\frac{1}{\sqrt2}r_3  & r_1 & 0 & 0\\
		r_1 & \frac{1}{\sqrt2}r_3   & 0 & 0\\
		0 & 0 & -\frac{1}{\sqrt2}r_3   & r_2\\
		0 & 0 & r_2 & -\frac{1}{\sqrt2}r_3 
	\end{pmatrix}
	~.
}
The naive (i.e.\ excluding accidental symmetry enhancements) symmetry breaking pattern,
\equ{
	\label{TwoMus:NaiveSymmetryBreaking}
	\SU4 ~\xrightarrow{\text{explicit}}~ \U1^3  ~\xrightarrow{\text{spontaneous}}~ \U1
	~,
}
dictates the structure of the coset.
For those Goldstone fields corresponding to exact symmetries of the action \eqref{LSM:ExplicitSymmetryLagrangian} we have
\equ{
	\label{LSM:TwoMus:GoldstoneCoset}
	U_G = \exp\left\lbrace\sfrac i2\left(\mu_1 t + \frac{\chi_1}{v\cos\frac\vartheta2}\right)\diag(1,-1,0,0)
	+
	\sfrac i2\left(\mu_2 t + \frac{\chi_2}{v\sin\frac\vartheta2}\right)\diag(0,0,1,-1)
	\right\rbrace
	~.
}
The chemical potentials $\mu_1$ and $\mu_2$ are determined by the classical solution \eqref{LSM:ChemicalPotentials}. The radial amplitude $v$ and the angle $\vartheta$ are fixed by the general Noether-matrix condition \eqref{LSM:TwoMus:NoetherCondition}.
The coset factor for the spectator modes
can be written up to $\order{1/v}$ re-orderings as 
\fontsize{10.5pt}{\baselineskip}
\begin{align}
\label{LSM:TwoMus:SpectatorCoset}
&U_\varphi 
= 
\exp\!\sfrac iv
\left\lbrace\!\!
\left(\!
\arraycolsep=1.7pt\def\arraystretch{1}
\begin{array}{cccc}
0 & 0 & \varphi_1-i \varphi_3 + (\varphi_2 -i \varphi_4) \tan \vartheta & 0
\\
0 & 0 & 0  & \frac{1}{\cos \vartheta }(\varphi_2-i \varphi_4)
\\
\varphi_1+i \varphi_3 +(\varphi_2 +i \varphi_4) \tan \vartheta  & 0  & 0 & 0
\\
0 & \frac{1}{\cos \vartheta }(\varphi_2+i \varphi_4) & 0 & 0 
\end{array}
\!\right)
\right.
\nonumber
\\[1ex]
&
+\!
\left.
\left(\!
\arraycolsep=1.7pt\def\arraystretch{1}
\begin{array}{cccc}
0 & -\frac{i}{2 \cos\vartheta/2} \varphi_9 & 0 & \frac{1}{\cos\vartheta}(\varphi_6-i \varphi_8)
\\
\frac{i}{2 \cos\vartheta/2} \varphi_9 & 0 & \varphi_5-i \varphi_7 + (\varphi_6 -i \varphi_8) \tan \vartheta  & 0
\\
0  & \varphi_5+i \varphi_7 + (\varphi_6 +i \varphi_8) \tan \vartheta  & 0 & -\frac{i}{2 \sin\vartheta/2}\varphi_{10}
\\
\frac{1}{\cos \vartheta}(\varphi_6+i \varphi_8) &0 & \frac{i}{2 \sin\vartheta/2}\varphi_{10} & 0 
\end{array}
\!\right)
\!\!\right\rbrace\!,
\end{align}
\normalsize
such that the quadratic kinetic term  in Eq.\ \eqref{LSM:ExplicitSymmetryLagrangian} is conveniently diagonal in $\varphi_i$.
Due to accidental enhancements for certain charge configurations there appear more massless modes coming from $U_\varphi$. 

With the data specifying the coset construction 
at hand, we proceed to expand the Lagrangian in $r_\ga,\varphi_a$ and $\chi_i$ as instructed by Eq.\ \eqref{LSM:LagrangianExpansion_Formal}.  
In the spirit of Section~\ref{ssc:SymmetryDispersions} we read off the tree-level propagators from the quadratic piece $\mathcal{L}^{(2)}$.
%
The fluctuating fields are always ordered such that $D^{-1}(k)$ optimally takes block-diagonal form:
\equ{
	\label{LSM:TwoMus:FullInversePropagtor}
	D^{-1}(k) = 
	\diag \left(D^{-1}(k)\vert_{r_1,r_2,\chi_1,\chi_2,r_3}\,,\,D^{-1}(k)\vert_{\varphi_i,\,i=1,..,4}\,,\,D^{-1}(k)\vert_{\varphi_i,\,i=5,..,8} \,,\,D^{-1}(k)\vert_{\varphi_i,\,i=9,10}\right)
	~.
}
The explicit expressions for the various blocks are given in the second paragraph of Appendix~\ref{app:Propagators}.
Solving Eq.\ \eqref{LSM:RootsOfInversePropagator}
for the first block 
we find three massive radial modes with masses of the order
\equ{
	M^{(i)}_r \sim \, m^{(i)}_r(Q_2/Q_1) \,\sqrt Q_1
	~,~~i=1,2
	\,\qand\,
	M^{(3)}_r \sim \, m^{(3)}_r(Q_2/Q_1,g_3) \,\sqrt Q_1
	~, 
}
where it suffices\footnote{
	Since these massive modes do not appear in the low-energy physics, we do not care about the precise dependence of their mass on $\gl_1,\gl_2, g_3$ and the ratio $Q_2/Q_1$.} to note that the functions $m_r^{(i)}>0$ within $16\gl_1>3\gl_2>0$ and suitably adjusted $g_3$.
In addition to the massive modes, we obtain two relativistic Goldstones,
\equ{
	\label{LSM:GenericCase:Goldstones}
	\omega_\chi^{(i)} = c_s^{(i)} \abs{\textbf k}  + \order{v^{-2}}
	\quads{,} i=1,2~,
}
with speed of sound
\equ{
	\label{LSM:GenericCase:SpeedsOfSound}
	c_s^{(1)} = \frac{1}{\sqrt2}
	\qand
	c_s^{(2)} = \frac{\lambda _2 \left(\cos^2\vartheta-1\right) \left(2 \lambda_2\, \cos^2\vartheta -16 \lambda_1 +\lambda_2\right)}{8 \lambda _2^2\, \cos^4\vartheta + \left(32 \lambda _1 \lambda _2-22 \lambda_2^2\right) \cos^2\vartheta + 256 \lambda _1^2-\lambda _2^2}
	~,
}
where $\vartheta$
is formally the solution to Eq.\ \eqref{LSM:SU4ChargesRelation} which cannot be given in a closed form. Anyhow, 
the speed of sound of the second relativistic Goldstone has 
to be determined via non-perturbative methods for each ratio $Q_2/Q_1$.

Contrary to the first Goldstone mode, which exhibits the by now familiar universal dispersion relation,
the speed of sound of the second gapless mode depends on the specifics of the physical system, i.e.\ the Wilsonian parameters $\gl_1,\gl_2, c_{nm}$, as well as on the ratio $Q_2/Q_1$. Inside the admissible region $16\gl_1>3\gl_2>0$, it is $c_s^{(2)}<1$ as required by causality.
Once $\gl_2=0$, also $c_s^{(2)}=0$ and we thus recover the predictions of the Wilson-Fisher-like fixed point discussed in Section~\ref{sc:OneMu}.

In a completely analogous fashion we analyze the dispersions from the spectator part $\left.D^{-1} (k) \right\vert_{\varphi_i}$. 
The two $4\times4$ blocks give  eight generically massive modes with pairwise the same mass,
\equ{
	\label{LSM:TwoMus:SpectatorMasses}
	\underbrace{M_\varphi^{(i)} \sim  \,m_{\varphi}^{(i)}(Q_2/Q_1)\,\sqrt{Q_1} 
	}_{2 \text{ times}}
	\quads{,}i=1,...,4~.
}
Again, the mass parameters $m_{\varphi}^{(i)}$ are always strictly positive for $i=2,3,4$ within the allowed parameter range. For generic $Q_2/Q_1$ also $m_{\varphi}^{(1)}$ is non-zero.
In addition, there are two diagonal massive modes, $\varphi_9$ and $\varphi_{10}$, with masses $\mu_1$ and $\mu_2$, respectively.
Based on the spectrum we have just derived from propagator \eqref{LSM:TwoMus:FullInversePropagtor} and using energy formula \eqref{LSM:TwoMus:CondensateEnergy} and the speeds of sound in Eq.\ \eqref{LSM:GenericCase:SpeedsOfSound}, we arrive at the following expression for the anomalous dimension
of the lowest scalar operator with generic charges $Q_1>Q_2>0$:
\begin{align}
\label{LSM:GenericRhos:AnomalousDimension}
\Delta(Q_1,Q_2)=&\,\,
c_{3/2}\left(Q_2/Q_1\right) \left(\frac{Q_1}{4\pi}\right)^{3/2} + c_{1/2}\left(Q_2/Q_1\right) \left(\frac{Q_1}{4\pi}\right)^{1/2}
\\[1ex]
&\,\,
-0.0937
-0.1325
\, c_s^{(2)}\left(Q_2/Q_1\right) + \order{Q_1^{-1/2}}
~.
\nonumber
\end{align}
Again, the asymptotic expansion in $1/Q_1\ll1$ is of the general form \eqref{LSM:AnomalousDimension} with $f_P=0.1325
\, c_s^{(2)}$. 
In addition to the ratio $Q_2/Q_1$, the ignorance coefficients and the model-dependent speed of sound depend on the Wilsonian parameters of the effective theory.
This  expansion and the associated spectrum 
exhibit two interesting limiting cases.

\paragraph{Case I.}
First of all, as can be deduced by slightly moving away from the extreme case analyzed in Section~\ref{ssc:TwoMus_OneU1}, 
taking $Q_1 \approx Q_2$ such that (for reasonable values of $c_{nm}$ in Eq.\ \eqref{LSM:TwoMus:Bmatrix}) $b_1 \approx b_2$ and hence
\equ{
	\cos\vartheta = \frac{16\gl_1 - \gl_2}{2 \left(16\gl_1 - \gl_2\right)} \frac{Q_1-Q_2}{Q_1} + \order{Q_1^{-2}} \,\ll\,1
	~,
}
results in a symmetry enhancement.
In detail, 
the first spectator mass parameter in Eq.\ \eqref{LSM:TwoMus:SpectatorMasses} becomes subleading, $m_{\varphi}^{(1)} \sim \order{1/\sqrt{Q_1}}$, 
resulting in the appearance of 
two additional relativistic 
Goldstones with the same model-dependent  speed of sound $c_s^{(2)}$.
Precisely at $Q_2=Q_1$, when the coset parametrization in Eq.\ \eqref{LSM:TwoMus:SpectatorCoset} becomes singular ($\cos\vartheta=0$), the accidental symmetry is enhanced to a true symmetry of the action \eqref{LSM:ExplicitSymmetryLagrangian} and we recover the spectrum of 
Table~\ref{tb:ToyModel:Spectrum}.

The associated expansion for the anomalous dimension of an operator with charges $Q_1\approx Q_2$ was computed in Eq.\ \eqref{LSM:OneRho:AnomalousDimension}.
We stress the clear order-one difference of that expansion compared to Eq.~\eqref{LSM:GenericRhos:AnomalousDimension} where $Q_1\neq Q_2$.
In the limiting case $Q_2\rightarrow Q_1$, there is a factor of 3 in front of the vacuum energy contribution associated with the non-universal speed of sound. Since $c_s^{(2)}$ is perturbatively undetermined but nevertheless bounded by causality, one cannot reabsorb this factor by any redefinition. Hence, we have a sharp way to distinguish $Q_1\approx Q_2$ from $Q_1 \neq Q_2$, already at the analytic level.

\paragraph{Case II.}
On the other side, we can take the opposite limit to fix a large hierarchy among the two charge scales by choosing $Q_2 \ll Q_1$. In that case, 
\equ{
	\cos\vartheta = 1 - 2 \sqrt{\frac{16 \lambda_1 + 5 \lambda_2}{16 \lambda_1 - 3 \lambda_2 } }\, \frac{Q_2}{Q_1} \,+\, \order{Q_1^{-3/2}} \,\gg\,0
} 
so that $b_2\approx 0$.
All spectator modes 
in Eq.\ \eqref{LSM:TwoMus:SpectatorMasses} have large masses at large $Q_1$. In addition, the speed of sound in Eq.\ \eqref{LSM:GenericCase:SpeedsOfSound} of the model-dependent Goldstone becomes suppressed by $Q_1$ according to
\equ{
	\omega_\chi^{(2)} = \frac{2\sqrt{2\gl_2}}{\sqrt[4]{(16\gl_1-3\gl_2)(16\gl_1+5\gl_2)} } \,\sqrt{\frac{\rho_2/b_2}{\rho_1/b_1}}\, \abs{\textbf k}  + \order{1/\rho_1}
	~.
} 
Precisely when $\rho_2=0$ and $\cos\vartheta=1$, there is a symmetry restoration\footnote{For $\rho_2=0$ 
a larger symmetry enhancement is possible for the values $\mu_1=\mu_2$ in
    Eq.~\eqref{LSM:TwoMus:SU4ClassicalPhi} and $\vartheta=0$ in Eq.~\eqref{LSM:TwoMus:EigenvalueMatrixSU4}. In that case, the coset is given by $\U3/\U2$ resulting into one relativistic and two non-relativistic Goldstone fields.
}
of the form
\equ{
	\label{Case2:SymmetryBreakingB}
	\SU4 ~\xrightarrow{\text{explicit}}~ 
	\U3 ~\xrightarrow{\text{spontaneous}}~ \U2
	~,
} 
which promotes $\omega_\chi^{(2)}$ from Eq.~\eqref{LSM:GenericCase:Goldstones} into a non-relativistic Goldstone with  quadratic (Galilean) dispersion relation,
\equ{
	\omega_\chi^{(2)} = \frac{\sqrt[4]{16\gl_1+5\gl_2}}{\sqrt{16\gl_1-3\gl_2}} \sqrt{\frac{1}{\rho_1/b_1}}\abs{\textbf k}^2 + \order{1/\rho_1}
	~.
} 
This fact is in accordance with the preceding literature finding that having only one non-vanishing $\U1$ charge results into one relativistic and at most  a bounce of non-relativistic Goldstones.

In total, we see that the leading energy on $S^2$ when $Q_1 \gg Q_2 \geq 0$ becomes 
\begin{align}
\label{LSM:TwoRhos:AnomalousDimension}
\Delta(Q_1,Q_2) =&\,\, \frac{\left(16 \lambda_1+5\lambda_2\right)^{1/4}}{3} \left(\frac{Q_1}{4\pi b_1}\right)^{3/2}
+ \frac{
	1
}{4 \left(16 \lambda_1+5\lambda_2\right)^{1/4}}  
\left(\frac{Q_1}{4\pi b_1}\right)^{1/2} 
-0.0937
+\order{Q_1^{-1/2} }
.
\end{align}
This formula has to be especially compared with the energy expansion of the opposite liming scenario $Q_1 \sim Q_2$ in Eq.~\eqref{LSM:OneRho:AnomalousDimension}.
Most importantly, 
there exists no model-dependent contribution to order $Q_1^0$, as it is either suppressed by $1/ Q_1$ when $0<Q_2\ll Q_1$ or it is strictly zero in case $Q_2=0$ (recall that a non-relativistic Goldstone has by definition vanishing vacuum energy).
Therefore, for this charge configuration there is no qualitatively tractable difference up to order one in the large-charge expansion compared to the prediction at a Wilson-Fisher-type fixed point.
\paragraph{}
All in all, the detailed analysis of the dispersion relations reveals the origin of the order-one terms in Eq.\ \eqref{LSM:AnomalousDimension}, as
summarized in Table~\ref{tb:GenericFP:Summary}.
By the semi-classical analysis we have verified the existence of the universal Goldstone with speed of sound $1/\sqrt2$ for any charge configuration.
Most crucially, we have seen that  the model-dependent contribution to $f_P$ for $P=2$ 
exhibits three qualitatively distinct regions (last column in Table~\ref{tb:GenericFP:Summary}),  depending on the charge configuration at the multi-charge fixed point.

%
%

\begin{table}[t]
	\renewcommand*{\arraystretch}{1.1}
	\centering
	\begin{tabular}{|l|l||c|r|c|c|}
		\multicolumn{2}{c}{}&\multicolumn{4}{c}{\textbf{
				gapless modes at the multi-charge fixed point}}
		\\[.2ex] \hline
		charge setup	&	effective & universal &model-dependent & coset &
		\\ 
		with $Q_1 \gg 1$
		& parameters  & $c_s=1/\sqrt2$ 
		&  gapless modes 
		& space & $f_2$  
		\\
		\hline\hline
		\rule{0pt}{3.5ex}
		$Q_2\, \approx\, Q_1$ & $b_2\,\approx \,b_1$ & 
		yes & 3 relativistic with $c_s^{(2)}$ 
		& $\frac{\U2 \times \SU2}{\vphantom{\bar f}\SU2'}$ 
		& $3\times 0.13\, c_s$
		%
		%
		\\[.8ex]
		\hline
		\rule{0pt}{3ex}
		$Q_2<Q_1$  & $b_2<b_1$ & yes
		& 1 relativistic with $c_s^{(2)}$ & \multirow{2}{*}{$\frac{\U1^3}{\vphantom{\bar f}\U1'}$}
		& $0.13\, c_s$
		\\
		\cline{1-4}\cline{6-6}
		\rule{0pt}{3.5ex}
		$Q_2\ll Q_1$  & $b_2\approx0$ & yes & 1 relativistic with $c_s^{(2)}\ll1$ & & \multirow{2}{*}{0}
		\\
		\cline{1-5}
		\rule{0pt}{3.5ex}
		$Q_2=0$ & $b_2=0$ & yes & 1 non-relativistic & $\frac{\U2}{\vphantom{\bar f}\U1}$ 
		&
		\\[.8ex]
		\hline
	\end{tabular}
	\renewcommand*{\arraystretch}{1}
	\caption{The table summarizes the light spectrum 
		(including accidental enhancements) 
		supported by 
		various charge configurations
		at the multi-charge fixed point. 
		The second column refers to the coefficients defined in Eq.~\eqref{LSM:TwoMus:Bmatrix}.
		The third column stresses the existence of a universal relativistic Goldstone for any charge configuration, while
		the forth column specifies the 
		non-universal Goldstone modes 
		that appear.
		In the fifth column, we also provide  
		the associated coset space. 
		The last column gives the contribution of the non-universal, model-dependent light spectrum to formula \eqref{LSM:AnomalousDimension}.
	}\label{tb:GenericFP:Summary}
\end{table}

\subsection{Loop suppression}
\label{ssc:LoopSuppression}

The stability of the large-charge construction under quantum corrections has been verified in \cite{Hellerman:2015nra} for a pure $\U1$ theory 
and in \cite{Alvarez-Gaume:2016vff} for any $O(n)$ vector model 
when the light spectrum includes a universal relativistic Goldstone plus additional non-relativistic fields.
For models with similar characteristics we refer to those previous papers.
Instead, we demonstrate how the suppression of quantum corrections works at large  $Q_1$ in the novel situation with multiple relativistic Goldstones.
This shall be done using the path integral formulation \cite{Loukas:2016ckj} by integrating out any massive modes 
while treating the light \textsc{dofs} as background fields.

By the semi-classical analysis in the previous paragraphs we have found multiple massive modes. 
Their masses scale with $\mu\sim\order{\sqrt{Q_1}}$.
Therefore, any $\mu$--massive mode can be safely integrated out at $Q_1\gg1$, as its loops will be suppressed by inverse powers of the large parameter.
This also means that any higher term of such massive mode coupled to the light \textsc{dof}s 
is irrelevant for the leading low-energy action.
Consequently, the interesting for us dispersion relations of the Goldstone sector in our theory are determined just by setting all $\mu$-massive modes to the minimum of their respective scalar potential.

We demonstrate the suppression 
using the charge configuration of Section~\ref{ssc:TwoMus_OneU1}.
The more generic situation with two relativistic Goldstones works in a similar fashion. 
After the spectrum has been determined
by analyzing the quantum Lagrangian $\mathcal L^{(2)}$ in Eq.\ \eqref{LSM:OneRho:QuadraticLagrangian},
we have to show that the contribution of higher terms $\mathcal L^{(m\geq3)}$ comes only sub-leading to the previously derived dispersion relations
by integrating out all massive modes, $r_\ga$ and $\varphi_a$.
However, from the form of $\mathcal L^{(2)}$ 
it becomes clear that some of the massive fields are coupled to the light \textsc{dof}s $\chi_i$ for $i=1,2,3,4$. 
Therefore, we  need to explicitly diagonalize 
the quadratic Lagrangian, instead of just looking at the roots of $\det D^{-1}(k)$, as we did above. For this purpose, we define two new Goldstone fields 
\equ{
	\chi_\pm = \left({\chi_2 \pm \chi_1}\right)/{\sqrt2}
	~.
}
Next, for the first four radial modes we determine the non-trivial minimum of their scalar potential: 
\equ{
	\begin{align}
	r_1^\text{min} = \frac{2 \lambda _2\, \dot{\chi }_+ - \left(16 \lambda_1 - \lambda_2\right) \dot{\chi }_-}{4 \sqrt{2} \lambda _2\, \mu }
	+ \order{\mu^{-3}}
	&\qand
	r_2^\text{min} = 
	\frac{2 \lambda _2\, \dot{\chi }_+ + \left(16 \lambda_1 - \lambda_2\right) \dot{\chi }_-}{4 \sqrt{2} \lambda _2\, \mu }
	+ \order{\mu^{-3}}
	\nonumber
	\\[1ex]
	r_3^\text{min} = 
	\frac{\left(16 \lambda _1-\lambda _2\right) }{4 \lambda _2 \mu }\, \dot{\chi}_3 + \order{\mu^{-3}}
	&\qand 
	r_4^\text{min} = \frac{\left(16 \lambda _1-\lambda _2\right) }{4 \lambda _2 \mu }\, \dot{\chi}_4 + \order{\mu^{-3}}
	~,
	\end{align}
}
where the $\chi$--Goldstones are treated as background fields.
All other $\mu$-massive modes have a trivial minimum at the origin, 
$r_\ga^\text{min}=0, ~\ga=5,6,7$ and $\varphi_a^\text{min} = 0, ~\forall a$\,.
We are now in a position to perform the path integral over the massive radial $r_\ga$ and spectator $\varphi_a$ modes,
\equ{
	i\tilde S[\chi] =  \log \int \mathcal D r \mathcal D \varphi \, \e^{iS[r_\ga,\varphi_a,\chi]}
	~,
}
in order to read off the resulting action in the Goldstone fields:
%
\begin{align}
\tilde{\mathcal L} = &\,\,
\frac{4 \mu ^3}{3 \sqrt{16 \lambda _1-\lambda _2}}
-\frac{\mu \mathcal{R}}{4 \sqrt{16 \lambda_1-\lambda_2}}
+\frac{2 \mu ^{3/2}}{\sqrt[4]{16 \lambda _1-\lambda _2}}\, \dot{\chi }_+
\\[1ex]
&\,\,
+\dot{\chi}_+^2-\frac{1}{2} \left(\nabla \chi_+\right)^2
+\frac{\left(16 \lambda _1+\lambda _2\right)}{4 \lambda_2} \dot{\chi}_-^2
-\frac{1}{2} \left(\nabla \chi_-\right)^2
\nonumber
\\[1ex]
&\,\,
+\frac{\left(16 \lambda_1+\lambda_2\right)}{4 \lambda_2}\,\dot\chi_3^2
-\frac{1}{2} \left(\nabla \chi_3\right)^2
+\frac{\left(16 \lambda_1+\lambda_2\right)}{4 \lambda_2}\, \dot\chi_4^2
-\frac{1}{2} \left(\nabla \chi_4\right)^2
~+~ 
\order{\mu^{-1/2}}
~.
\nonumber
\end{align}
As expected, integrating out the massive modes reproduces the low-energy spectrum found semi-classically in Section~\ref{ssc:TwoMus_OneU1}
with quantum corrections due to massive fluctuations being of higher orders in $1/\mu$.
This ensures the stability of the leading Goldstone dispersion relations stated in Table~\ref{tb:ToyModel:Spectrum}. 

\subsection{Generalizing to $\SU N$ theory}
\label{ssc:MultiChargeFixedPoint_SUN}

Let us generalize the previous results at the multi-charge fixed point (for generic $g_i$ in Eq.\ \eqref{LSM:ScalarPotential}) to $\SU{2k}$ matrix theory. The homogeneous  solution to the classical \textsc{eom}s can be brought to block-diagonal form,
\equ{
	\label{LSM:SUN_GenericFP:ClassicalSolution}
	\Phi(t) = \frac{1}{\sqrt2}
	\renewcommand*{\arraystretch}{1.2}
	\left(
	\begin{array}{ccccccc}
		\cline{1-2}
		\multicolumn{1}{|c}{0} & \multicolumn{1}{c|}{v_1\,\e^{i\mu_1 t}}  &  & \\
		\multicolumn{1}{|c}{v_1\,\e^{-i\mu_1 t}} & \multicolumn{1}{c|}{0} &  &  \\
		\cline{1-2}
		\cline{3-4}
		&   & \multicolumn{1}{|c}{0} &  \multicolumn{1}{c|}{v_2\,\e^{i\mu_2 t}} \\
		&  & \multicolumn{1}{|c}{v_2 \,\e^{-i\mu_2 t}}  &  \multicolumn{1}{c|}{0} \\
		\cline{3-4}
		& & & & \ddots
		\\
		\cline{6-7}
		& & & & & \multicolumn{1}{|c}{0}&\multicolumn{1}{c|}{v_k\,\e^{i\mu_k t}}\\
		& & & & & \multicolumn{1}{|c}{v_k \,\e^{-i\mu_k t}} &\multicolumn{1}{c|}{0}\\
		\cline{6-7}
	\end{array}
	\right)
	~,
}
with the associated finite Noether-current matrix being diagonal,
\equ{
	\label{NoetherCurrent_SU2k}
	J_0 = \diag \left(\mu_1 v_1^2\,,\, - \mu_1 v_1^2 \,,\, ... \,,\, \mu_k v_k^2\,,\, - \mu_k v_k^2 \right) ~\in~ \su{2k}
	~.
}
In the case of the $\SU{2k+1}$ matrix model the classical solution has again $k$ such blocks and zero elsewhere. The Noether-current matrix is similarly modified to
\equ{
	\label{NoetherCurrent_SU2k+1}
	J_0 = \diag \left(\mu_1 v_1^2\,,\, - \mu_1 v_1^2\,,\, ... \,,\, \mu_k v_k^2\,,\, - \mu_k v_k^2\,,\, 0\right) ~\in~ \su{2k+1}
	~.
}
It is convenient to introduce generalized polar coordinates
\begin{align}
\label{GeneralizedPolarCoordinates}
v_1 = v\cos\vartheta_1
~,~
v_2 = v\sin\vartheta_1\cos\vartheta_2
~,~...~,~
v_{k-1} =&\, v\sin\vartheta_1\cdots \sin \vartheta_{k-2}\cos\vartheta_{k-1}
~,
\nonumber
\\
v_{k} =&\, v\sin\vartheta_1\cdots \sin \vartheta_{k-2} \sin\vartheta_{k-1}
~,
\end{align}
to parametrize the radial \textsc{vev}s $v_i$\,.
Given an $\SU N$ matrix model with $N=2k$ or $N=2k+1$, the classical \textsc{eom}s can then be schematically expressed as 
\equ{
	\mu_j = v^2 \,f_j \left(g_1,g_2,g_4\,;\,\vartheta_1,...,\vartheta_{k-1}\right)
	\,+\,
	\text{subleading $\mathcal R$-dependent terms,}
	\quads{\text{with }}
	j=1,...,k  
	~.
}
The $k$ chemical potentials $\mu_j$ are thus functionally determined by the Wilsonian couplings and the polar angles, while their scaling is generically of order $v^2$.
From Eq.\ \eqref{NoetherCurrent_SU2k} and \eqref{NoetherCurrent_SU2k+1} we see that it is possible to fix 
up to $k=\floor{N/2}$ 
global $\U1$ charges $Q_j$.
Taking $Q_1\gg1$  it is clear that  $v^4\sim\order{Q_1}$ so that we can write in the spirit of Eq.\ \eqref{LSM:AnomalousDimension} an asymptotic expansion of the anomalous dimension:
\equ{
	\label{SUN:AnomalousDimension}
	\Delta(Q_1,...,Q_k)	=
	c_{3/2}(Q_i/Q_j) \left(\frac{Q_1}{4\pi}\right)^{3/2} + c_{1/2}(Q_i/Q_j) \left(\frac{Q_1}{4\pi}\right)^{1/2} -0.0937
	- f_2(Q_i/Q_j) + \order{Q_1^{-1/2}}
	.
}
The ignorance coefficients $c_{3/2}$, $c_{1/2}$ and the non-universal order-one contribution $f_2$ depend on the ratio $Q_i/Q_j$ for $i<j$ as well as on the Wilsonian parameters $g_1,g_2$ and $g_4$.

Inspecting the form of the classical solution \eqref{LSM:SUN_GenericFP:ClassicalSolution} it is possible, by readily generalizing Eq.\ \eqref{LSM:TwoMus:EigenvalueMatrixSU4} and \eqref{LSM:TwoMus:TrafoMatrixSU4},
to bring $\Phi(t)$ to the form \eqref{LSM:ClassicalSolutionSCHEMA} 
where $\Phi_0$ represents the time-independent block-matrix and the direction of the time-dependent \textsc{vev} is given by
\equ{
	\sum_{j=1}^{\floor{N/2}}\mu_j\, h^j = 
	\left\lbrace
	\renewcommand*{\arraystretch}{1.2}
	\begin{array}{ll}
		\diag	\left(\mu_1\,,\,-\mu_1\,,\, ...\,,\, \mu_k\,,\,-\mu_k\right) & \text{ for }~ N=2k\\
		\diag	\left(\mu_1\,,\,-\mu_1\,,\, ...\,,\, \mu_k\,,\,-\mu_k\,,\,0\right) &  \text{ for }~ N=2k+1
	\end{array}
	\right.
	~.
}
From here we read off the symmetry breaking pattern for generic $Q_j \neq0$ for all $j=1,...,k$ (implying generic polar angles in Eq.\ \eqref{GeneralizedPolarCoordinates}), as explained in Section~\ref{ssc:SymmetryDispersions}:
\equ{
	\label{GenericFP:SymmetryBreakingSUN}
	\begin{array}{lclcl}
		\SU{2k} &\xrightarrow{\text{explicit}}~&  \U1^{2k-1}& \xrightarrow{\text{spontaneous}}&  \U1^{k-1}
		\\[1ex]
		\SU{2k+1} &\xrightarrow{\text{explicit}}~ & \U1^{2k} &\xrightarrow{\text{spontaneous}} & \U1^{k}
		\,~.
	\end{array}
}
Thus, we expect in both cases $k=\floor{N/2}$  relativistic Goldstone fields. One of these  modes should always possess the universal dispersion relation with $c_s^{(1)} = 1/\sqrt2$ and the rest 
will have speed of sounds $c_s^{(j)}$ for $j=2,...,k$, depending on the microscopic details of the theory and the precise charge configuration.
Consequently, we generically expect $k-1$ ignorance parameters in the energy expansion  
\eqref{SUN:AnomalousDimension} at order one.

\section{The Wilson-Fisher-like fixed point}
\label{sc:OneMu}

The second branch of the homogeneous 
solution to the \textsc{eom}s \eqref{LSM:EOMs} with one chemical potential appears 
inside a special region in the space of Wilsonian parameters, 
when $g_1=g_4=0$ in the  potential \eqref{LSM:ScalarPotential}.

\subsection{The classical solution}

The classical solution of $\SU N$ matrix theory with one chemical potential $\upmu$ 
can be written as 
\begin{align}
\label{OneMu:GeneralClassicalSolution}
\Phi(t) =&\,\, 
\sfrac{1}{\sqrt2}
\begin{pmatrix}
0   & \upupsilon_1 \e^{i\upmu t} & 0       & \cdots & 0  & 0      
\\
\upupsilon_1 \e^{-i\upmu t} & 0   & \upupsilon_2 \e^{i\upmu t}       & \cdots & 0    & 0   
\\
0   & \upupsilon_2 \e^{-i\upmu t} & 0     & \cdots& 0   & 0
\\
\vdots&\vdots&\vdots  &     & \vdots  & \vdots \\
0   & 0   & 0   & \cdots & 0       & \upupsilon_{N-1}\e^{i\upmu t}
\\
0  & 0   & 0   & \cdots & \upupsilon_{N-1} \e^{-i\upmu t} & 0
\end{pmatrix}
\\[1ex]
=&\,\,
\sum_{i=1}^{N-1} \frac{\upupsilon_i}{\sqrt2} \, \left(\e^{i\upmu t}\,E^{\alpha_i} + \e^{-i\upmu t}\, E^{-\alpha_i} \right)
=
\sum_{i=1}^{N-1} \frac{\upupsilon_i}{\sqrt2} \, \Ad[\e^{i\upmu t\, h}]\left(E^{\alpha_i} +  E^{-\alpha_i} \right)
~,
\nonumber
\end{align}
with the direction of the time-dependent \vev in the language of Eq.\ \eqref{LSM:ClassicalSolutionSCHEMA} given by 
\equ{
	\label{LSM:OneMu:hDirection}
	h = 
	\left\lbrace
	\renewcommand*{\arraystretch}{1.2}
	\begin{array}{ll}
		\sfrac12\, \diag \left(2k-1,2k-3,...,1,-1,...,-2k+1\right)
		&\text{ for }~
		\SU{2k} 
		\\
		~~\,\diag \left(k,k-1,...,1,0,-1,...,-k\right)
		&\text{ for }~
		\SU{2k+1} 
	\end{array}
	\right.
	~.
	\renewcommand*{\arraystretch}{1}
}
$E^{\pm\alpha_i}$ are the ladder operators corresponding to the simple root $\ga_i$ for $i=1,...,N-1$.
The chemical potential is fixed by the \textsc{eom}s in terms of the curvature $\mathcal R$ and the Wilsonian coupling $g_2$ to
\equ{
	\label{LSM:OneMu:EOMs}
	\upmu= \sqrt{g_2\, \upsilon^4 + \frac{\mathcal R}{8}}
	\quads{\text{ where }}
	\upupsilon^2 = \sum_{i=1}^{N-1} v_i^2
	~.
}
Using generalized polar coordinates to parametrize the radial \textsc{vev}s $\upupsilon_i$ according to Eq.\ \eqref{GeneralizedPolarCoordinates},
the appearance only of the overall radius $\upupsilon$ in Eq.\ \eqref{LSM:OneMu:EOMs} 
shows the $O(N^2-1)$ symmetry of the classical ground state in the given branch.
In the chosen gauge the Noether matrix becomes diagonal: 
\begin{align}
J_0 
=&\,\,
\upmu\,\diag \left(\upupsilon_1^2,-\upupsilon_1^2+\upupsilon_2^2,...,-\upupsilon_{N-2}^2+\upupsilon_{N-1}^2,-\upupsilon_{N-1}^2\right)
\\[1ex]
=&\,\,
\upupsilon^4 \sqrt{g_2 + \frac{\mathcal R}{8\upupsilon^4}}\,\,
\diag \left(\cos^2\theta_1\,,\,-\cos^2\theta_1 + \sin^2\theta_1\cos^2\theta_2\,,\,...\right)
~,
\nonumber
\end{align}
From the first line we see that it suffices to take all $\upupsilon_i\geq0$.
Written in polar coordinates in the second line, the current density makes apparent that we can at most fix one independent large-charge scale.
Of course, this observation remains to be verified by the quantum analysis, where the appearance of a unique relativistic Goldstone is anticipated. 
The radial amplitude $\upupsilon$ carries the large scale, while the  polar angles $\theta_i$ parametrize the precise alignment of $J_0$ in the Cartan sub-algebra of $\SU N$. In other words, we can  take in terms of the  unique scale described by $Q\gg1$:
\equ{
	\Tr\, J_0^2 \sim \order{\upupsilon^8} \sim \order{Q^2}	
	~.
}

The low-energy physics and in particular the symmetry breaking pattern at large charge does not depend on the orientation of $J_0$ in the Cartan sub-algebra as specified by the $\theta$'s. 
In fact, once we consider the fluctuations on top of the large-charge vacuum, we see that only the massive modes depend on the polar angles. Since their masses scale with $\upmu\sim\order{\sqrt Q}$, they appear sub-leading in the large-charge expansion.
In contrast, the leading dispersion relations of the Goldstone fields (the ``good'' \textsc{dof}s in the low-enrgy regime) are independent of 
the precise orientation of $J_0\neq0$.

\paragraph{The Calogero-Moser system.}

Hence, it suffices to look at some selected charge configuration to outline the qualitative behavior of the system at this fixed point of the \textsc{rg} flow, while keeping notation condensed.
A particularly interesting setting arises when we orient our  current matrix along
\equ{
	\label{CalogeroMoser:ChargeCondition}
	J_0 \overset{!}{=}\, \frac{\rho}{b}\, \frac{2}{N^2-N}\, \,\diag\left(1,...,1,-(N-1)\right)
	~,
}
by choosing the amplitudes as
\equ{
	\label{CalogeroMoser:RadialAmplitudesEOMS}
	\upupsilon_j = \upupsilon\sqrt{\frac{2j}{N^2-N}}
	\quads{\text{for}}j=1,...,N-1
	~,
}
such that $\upmu\upupsilon^2=\rho/b$.
The angular momentum matrix \eqref{AngularMomentumInTheBody} takes then the characteristic form
\equ{
	K_{ij} = 
	\left\lbrace
	\begin{array}{cc}
		2/(N^2-N)\,\rho/b  & i\neq j\\
		0 & i=i
	\end{array}
	\right.
	~.
}
In terms of the (time-independent) eigenvalues of $\Phi(t)$ in Eq.\ \eqref{OneMu:GeneralClassicalSolution}, $a_i\equiv a_i(\upupsilon_j)$,
the homogeneous Hamiltonian simplifies  to 
\equ{
	\mathcal H = 
	\sfrac12 \sum_{i\neq j} \frac{(2/(N^2-N)\,\rho/b)^2}{(a_i-a_j)^2} + V(a_1,...,a_n)
	\quads{,}
	\sum_{i=1}^N a_i = 0
	~.
}
This Hamiltonian system describes the well-studied Calogero-Moser problem \cite{Polychronakos:2006nz},
namely $N$ identical particles on the real line, all with the same charge $2/(N^2-N)\,\rho/b$, repelling each other in a confining potential $V$.
From the form of the eigenvalues of $\Phi(t)$ on the classical solution (cf.\ Eq.\ \eqref{LSM:NecessaryConditionOnEigevalues}) (which is associated to the symmetry of $\mathcal H$ under reflection, $a_i\rightarrow -a_i$) we see that the configuration with the lowest energy is achieved, once the charged particles are aligned in mirror pairs around the origin. 
The fixed-charge condition \eqref{CalogeroMoser:ChargeCondition} determines the scaling of the radial amplitude to
\equ{
	\upupsilon = \left(\frac{1}{\sqrt{\gl_1}}\,\frac{\rho}{b}\right)^{1/4} \left(1 + \order{\rho^{-1}}\right)
	~,
}
so that the condensate energy can be expanded at large charge density $\rho$ as
\equ{
	\label{LSM:WilsonFisher:CondensateEnergyExpansion}
	E_0 = 
	\frac{\mu^2 \upupsilon^2}{2} + \frac{\mathcal R \upupsilon^2}{16} +\frac{\gl_1 \upupsilon^6}{6}
	=
	\frac23 \sqrt[4]{\lambda_1} \left(\frac{\rho}{b}\right)^{3/2}
	+
	\frac{\mathcal R}{16}
	\frac{1}{\sqrt[4]{\lambda_1}} \left(\frac{\rho}{b}\right)^{1/2}
	+ \order{\mathcal \rho^{-1/2}}
	~.
}
Here, we have used the effective coupling $\gl_1$ introduced in Eq.\ \eqref{LSM:EffectiveCouplings}.
This is special only to the $\su4$ algebra;
due to relation \eqref{SU4CasimirsRelation} it suffices to have $\gl_2=0$, i.e.\ $g_4=-\sfrac34g_1$.
Starting from the $\SU5$ matrix model one needs to set both $g_1=g_4=0$.

\subsection{Symmetry breaking and  spectrum}

As in the case of the generic fixed point, to analyze the structure of the Wilson-Fisher-like fixed point at large charge, we will concentrate on the sufficiently general
$\SU4$ matrix theory 
focusing on the particular Calogero-Moser configuration reviewed in the previous paragraph.
Around the classical ground state \eqref{OneMu:GeneralClassicalSolution} we write the fluctuations in the familiar form \eqref{LSM:CosetConstruction}
with the direction $h$ of the time-dependent \vev specified in the upper line of Eq.\ \eqref{LSM:OneMu:hDirection} for $k=2$ and the time-independent $\Phi_0$ obtained from \eqref{CalogeroMoser:RadialAmplitudesEOMS} for $N=4$.
The radial modes are determined by the condition $[\Phi_0,\Phi_\text{radial}]=0$ to
\equ{
	\Phi_\text{radial}=
	\renewcommand*{\arraystretch}{1.5}
	\left(
	\begin{array}{cccc}
		-\frac{r_1}{\sqrt{6}} & \frac{r_2}{2 \sqrt{3}}-\frac{r_3}{\sqrt{6}} & \frac{r_1}{2 \sqrt{3}} & \frac{r_3}{2} \\
		\frac{r_2}{2 \sqrt{3}}-\frac{r_3}{\sqrt{6}} & 0 & \frac{r_2}{\sqrt{6}}+\frac{r_3}{2 \sqrt{3}} & \frac{r_1}{2} \\
		\frac{r_1}{2 \sqrt{3}} & \frac{r_2}{\sqrt{6}}+\frac{r_3}{2 \sqrt{3}} & \frac{r_1}{\sqrt{6}} & \frac{r_2}{2} \\
		\frac{r_3}{2} & \frac{r_1}{2} & \frac{r_2}{2} & 0 \\
	\end{array}
	\right)
	\renewcommand*{\arraystretch}{1}
	~.
}
The coset for Goldstones corresponding to true symmetries of the action \eqref{LSM:ExplicitSymmetryLagrangian} is parametrized as
\begin{align}
\renewcommand*{\arraystretch}{1.2}	
U_G =
\exp \frac{i}{v}
\left(
\begin{array}{cccc}
\frac{\chi _1}{\sqrt{5}}+\frac{3 \chi _2}{\sqrt{5}}+\frac{3 \chi _3}{2} & 0 & 0 & 0 \\
0 & \frac{\chi _1}{\sqrt{5}}-\frac{2 \chi _2}{\sqrt{5}}+\frac{\chi _3}{2} & 0 & 0 \\
0 & 0 & -\frac{2 \chi _1}{\sqrt{5}}-\frac{\chi _2}{\sqrt{5}}-\frac{\chi _3}{2} & 0 \\
0 & 0 & 0 & -\frac{3 \chi _3}{2} \\
\end{array}
\right)
~,
\end{align}
while the naive coset factor for the spectator fields can be written as
\begin{align}
U_\varphi =
\fontsize{9.5pt}{\baselineskip}
\exp \frac{i}{v}\!
\left(\!\!\!\!
\arraycolsep=3pt\def\arraystretch{1.1}
\begin{array}{cccc}
-\varphi _2 & \sqrt{3} \varphi _1-i \varphi _4-\frac{i \varphi _6}{\sqrt{3}}+\frac{i \varphi _7}{\sqrt{6}}-\frac{i \varphi _9}{2} & \sqrt{2} \varphi
_2-i \varphi _5-\frac{i \varphi _8}{\sqrt{2}} & -\frac{3 i \varphi _7}{2} \\
\sqrt{3} \varphi _1+i \varphi _4+\frac{i \varphi _6}{\sqrt{3}}-\frac{i \varphi _7}{\sqrt{6}}+\frac{i \varphi _9}{2} & \varphi _2 & -i
\sqrt{\frac{2}{3}} \varphi _6-\frac{i \varphi _7}{2 \sqrt{3}}-\frac{i \varphi _9}{\sqrt{2}} & -i \sqrt{\frac{3}{2}} \varphi _8 \\
\sqrt{2} \varphi _2+i \varphi _5+\frac{i \varphi _8}{\sqrt{2}} & i \sqrt{\frac{2}{3}} \varphi _6+\frac{i \varphi _7}{2 \sqrt{3}}+\frac{i \varphi
	_9}{\sqrt{2}} & 0 & \sqrt{3} \varphi _3-\frac{i \sqrt{3} \varphi_9}{2}  \\
\frac{3 i \varphi _7}{2} & i \sqrt{\frac{3}{2}} \varphi _8 & \sqrt{3} \varphi_3+\frac{i \sqrt{3} \varphi_9}{2}  & 0 \\
\end{array}
\!\!\!\right)
\!\!.
\end{align}
\normalsize
Of course, this is one of the possible parametrizations for the coset space dictated by Eq.\ \eqref{LSM:CosetConstruction}.
Due to accidental enhancements at large charge some of the spectator fields become massless.
Implementing this particular realization we obtain the fluctuating Lagrangian (disregarding overall boundary terms):
\begin{align}
\mathcal L^{(2)} = &\,
\sfrac12\sum_{a=1}^9 \left(\dot{\varphi}_a^2 - (\nabla\varphi_a)^2 \right)
+\sfrac12\sum_{\ga=1}^3 \left(\dot{r}_\ga^2 - (\nabla r_\ga)^2 \right)
+\sfrac12\sum_{i=1}^3 \left(\dot{\chi_i}^2 - (\nabla\chi_i)^2 \right)
\\
&
-\upmu\left(\sqrt3+\frac{1}{\sqrt{3}}\right)\varphi_1 \dot\varphi_4
- 2 \sqrt{2} \upmu  \varphi_1 \dot\varphi_7
- 3\sqrt{2} \upmu\varphi_2 \dot\varphi_5
+\upmu\left(\sqrt3+\frac{1}{\sqrt{3}}\right) \varphi_3 \dot\varphi_6
- 2 \sqrt{\frac{2}{3}} \upmu \varphi_3 \dot\varphi_7 
\nonumber
\\
&
+ 2 \sqrt{\frac{2}{3}} \upmu r_1 \dot{\varphi}_1-2 \sqrt{2} \upmu r_1 \dot{\varphi}_3
+3 \sqrt{2} \upmu r_3 \dot{\varphi}_2
-\sqrt{\frac{2}{5}} \upmu\varphi_5 \dot{\chi}_1
+\frac{4\upmu}{\sqrt{5}}\varphi_8\dot{\chi}_1
\nonumber
\\
& 
+2 \sqrt{\frac{2}{5}} \upmu \varphi_5  \dot{\chi}_2 
+\frac{2\upmu}{\sqrt{5}}\varphi_8\dot{\chi}_2
+\sqrt{\frac{2}{5}} \upmu  r_3\dot{\chi}_1-2 \sqrt{\frac{2}{5}}\upmu r_3 \dot{\chi}_2
+2 \upmu  r_2 \dot{\chi}_3
\nonumber
\\
&
+2\sqrt{\frac{2}{3}}\upmu^2\varphi_4\varphi_7
- \frac{2}{3} \sqrt{2} \upmu ^2 \varphi_6\varphi_7 
+ \frac{2}{3} \sqrt{2} \upmu^2 r_1\varphi _4+2 \sqrt{\frac{2}{3}} \upmu^2 r_1 \varphi_6 
+ 4 \upmu^2 r_3\varphi_5
\nonumber
\\
&
- \left(\left(\frac{g_1+3g_3}{2 \lambda_1} -\frac{5}{6}\right)\upmu^2-\frac{g_1+3g_3}{16 \lambda_1} \mathcal R\right) r_1^2  -\left(2 \upmu^2 - \frac{\mathcal R}{4}\right) r_2^2 +2 \upmu^2 r_3^2
\nonumber
\\
&
+\frac{3}{2} \upmu ^2 \varphi_1^2+4 \upmu^2
\varphi_2^2+\frac{3}{2} \upmu ^2 \varphi_3^2+\frac{1}{6} \upmu ^2 \varphi _4^2+2 \upmu^2 \varphi _5^2+\frac{1}{6} \upmu ^2 \varphi_6^2+\frac{5}{6} \upmu^2 \varphi_7^2-\frac{1}{2} \upmu ^2 \varphi _9^2
\,\,+ \order{\upmu^{-1/2}}
~.
\nonumber
\end{align}

From here we read off the inverse propagator $D^{-1}(k)$. For the purposes of this section we are content to determine the roots of Eq.\ \eqref{LSM:RootsOfInversePropagator}. 
This gives us 12 massive modes,
\equ{
	\underbrace{\upomega_\upmu^{(1)} = \upmu + \order{\upmu^{-1}} }_{5 \text{ times}}
	\quads{,}
	\underbrace{\upomega_\upmu^{(2)} = 2\upmu + \order{\upmu^{-1}} }_{3 \text{ times}}
	\quads{,}
	\upomega_\upmu^{(3)} = 3\upmu + \order{\upmu^{-1}}
	\quads{,}
	\upomega_\upmu^{(4)} = 4\upmu + \order{\upmu^{-1}}
	\nonumber
	\\[1ex]
	\upomega_\mu^{(4)} = 2\sqrt2\upmu + \order{\upmu^{-1}}
	\qand
	\upomega_{r} = \sqrt{9 +\frac{g_1+3 g_3}{\lambda_1}}\,\upmu + \order{\upmu^{-1}}
	~,
}
as well as three Goldstone fields consisting of the universal relativistic and two non-relativistic  modes,
\equ{
	\upomega_{\chi}^{(1)} = \frac{\textbf k}{\sqrt2} + \order{\upmu^{-1}}
	\qand
	\underbrace{\upomega_{\chi}^{(2)} =  \frac{\textbf k^2}{2\upmu} + \order{\mu^{-2}} }_{ 2 \text{ times} }
	~.
}
Combining this light spectrum with the energy expansion \eqref{LSM:WilsonFisher:CondensateEnergyExpansion} we obtain the anomalous dimension formula at the Wilson-Fisher-like fixed point,
\equ{
	\label{LSM:WislonFisher:AnomalousDimension}
	\Delta(Q) = 
	\frac23 \sqrt[4]{\lambda_1} \left(\frac{Q}{4\pi b}\right)^{3/2}
	+
	\frac{1}{8\sqrt[4]{\lambda_1}} \left(\frac{Q}{4\pi b}\right)^{1/2}
	-0.0937
	+ \order{Q^{-1/2}}
	~.
}
As anticipated, it formally agrees with the asymptotic expansion at Wilson-Fisher fixed point (see Eq.\ (5.16) in \cite{Alvarez-Gaume:2016vff} as well as Eq.\ (2.56) in \cite{Loukas:2017lof}). 

The counting of the Goldstone \textsc{dofs}, $1+2\times2=5 = \dim \left(\U3/\U2\right)$, agrees with the symmetry breaking
\equ{
	\label{OneMu:SymmetryBreakingPattern}
	\SU4 ~\xrightarrow{\text{explicit}}~ 
	\U3 ~\xrightarrow{\text{spontaneous}}~ \U2 
	~.
}
This pattern 
generalizes the finding in \cite{Loukas:2017lof} concerning the $\SU3$ model to $\SU N$ matrix theory:
\equ{
	\label{OneMuSUN:SymmetryBreakingPattern}
	\SU N ~\xrightarrow{\text{explicit}}~ 
	\U{N-1} ~\xrightarrow{\text{spontaneous}}~ \U{N-2} 
	~.
}
Since only one  relativistic Goldstone, the universal $\omega_\chi^{(1)}$ is present in  this class of fixed points, the energy expansion on $S^2$ and the associated anomalous dimension are described by Eq.\ \eqref{Intro:SchematicEnergyExpansion} with $f_1=0$ and the same ignorance coefficients for any global $Q\gg1$.

\pagebreak
\section{Non-linear sigma models}
\label{sc:NonLinear}

For completeness we show how to directly write the non-linear sigma model on a general manifold, once the low-energy spectrum around the large-charge vacuum is known.
This serves as an important crosscheck for the previously derived asymptotic expansions of the anomalous dimension and is furthermore needed in order to compute fusion coefficients.
In the following we rederive the Goldstone spectrum for the special $Q_1=Q_2\equiv Q$ 
configuration and the general case with $Q_1\neq Q_2$.

\subsection{The $\U2$ coset}
\label{ssc:NonLinear_U2Coset}

In Section~\ref{ssc:TwoMus_OneU1} we have found that the generic fixed point can support in a certain limit a low-energy spectrum described by four relativistic \textsc{dof}s. 
This spectrum is dictated by the symmetry-breaking pattern of Eq.\ \eqref{LSM:ToyModel:SymmetryBreaking}
and the associated coset space turns out to be the $\U2$ Lie group.
Hence, we can parametrize our low-energy field variable $U\in \U2$ as
\equ{
	U(\chi;\pi_1,\pi_2,\pi_3)  = \e^{i\chi}\, \e^{i \sigma^3 \pi_3}\,\e^{i \sigma^2 \pi_1}\,\e^{i \sigma^3 \pi_2}
	\, \equiv\,  \e^{i\chi}\, \mathcal U(\pi_1,\pi_2,\pi_3)
	\quads{,}
	\mathcal U \in \SU2
	~,
}
where $\gs^i$ are the standard $2\times2$ Pauli-matrices and in the Euler-parametrization the angles are constrained within $\pi_1 \in[0,\frac{\pi}{2}],~\pi_2 \in [0,2\pi),~\pi_3 \in [0,\pi]$ and $\chi\in[0,2\pi)$.
The main building block of the non-linear action is given by
\equ{
	\label{NLSM:U2:BuildingBlock}
	\norm{\partial U} 
	\equiv
	\sqrt{\Tr\left(\partial_\mu U^\dagger \partial^\mu U\right)}
	=
	\sqrt{
		\abs{\partial\chi}^2+ \Tr\, \partial^\mu\mathcal U^\dagger\partial_\mu \mathcal U}
	\quads{\text{ where}}
	\abs{\partial\chi} \,\equiv\,\norm{\partial\chi} \equiv \sqrt{2\,\partial_\mu\chi\partial^\mu\chi}
	~.
}
Since our goal is to expand around a
charged 
vacuum where $\braket{U^\dagger \dot U} \sim \mu$ is large, the square root in $\norm{\partial U}$ is well-defined.

Following the analysis of the leading order terms 
outlined in \cite{Hellerman:2015nra}, 
we classify all possible scalar operators of dimension 3 
which are compatible with Lorentz and $\U2$ invariance.
We find that
the most general scale-invariant action in $2+1$ space-time dimensions 
can be written as 
\equ{
	\label{NLSM:U2Action}
	S = \int \d t \d\Sigma \left[
	\sfrac16\norm{\partial U}^3 - \frac{c_{\mathcal R}}{2} \mathcal R \norm{\partial U} \right] F(X,Y) + \order{\mu^{-1}}
	~,
}
where $\mathcal R$ is the Ricci scalar (e.g.\ of the two-sphere) and $c_{\mathcal R}$ an undetermined constant.
The functional freedom in writing the most general $\U2$-invariant action at fixed charge is encoded by $F(X,Y)$. At least up to order one in $\mu$ (the large \textsc{vev} of $U^\dagger \dot U$) this depends on  two dimensionless variables 
\equ{
	\label{NLS:DimensionlessVariables}
	X=\frac{\norm{\partial U}^2}{\abs{\partial\chi}^2} 
	\qand
	Y=\frac{\Tr \left(\partial_\mu U^\dagger \partial_\nu U\right)\Tr \left(\partial^\mu U^\dagger \partial^\nu U\right)}{\abs{\partial\chi}^4}
	~.
}
Using elementary Pauli-matrix algebra one can show that any other leading $\U2$- and Lorentz-invariant  combination of zero dimension can be expressed in terms of $X$ and $Y$.
There are further sub-leading, curvature-dependent quantities one can define
which we do not record here.

\paragraph{The classical equations of motion.}
Since we know from the analysis of the linear sigma model that our vacuum of lowest energy is homogeneous, we use that $\nabla U=0$ as an input in discussing  the \textsc{eom}s.
Notice that under this assumption $Y=X$.
By construction the theory \eqref{NLSM:U2Action} is invariant under $\SU2_L\times\SU2_R$ global transformations
implying the conserved Noether currents $J_L^\mu$ and $J_R^\mu$\,.
At the same time, $\U1$ invariance of the action implies the conserved current 
$J^\mu_{\U1}$ with zeroth component
\equ{
	\label{NLSM:U2:U1Current}
	J^0_{\U1} = \frac{\gd \mathcal L}{\gd \dot{\chi}} = 
	\left(\norm{\partial U} - \frac{c_{\mathcal R} \mathcal R}{\norm{\partial U}}\right)F(X,X)\, \dot{\chi} - \left( \sfrac16\norm{\partial U}^3 - \frac{c_{\mathcal R}}{2} \mathcal R \norm{\partial U} \right)\frac{\Tr \left(\dot{\mathcal U}^\dagger \dot{\mathcal U}\right)}{\dot{\chi}^3} \partial_X F(X,X)
	~.
}
For generic functional $F$ 
the classical \textsc{eom}s can be summarized by 
\equ{
	\frac{\d}{\d t} \left(\mathcal U^\dagger \dot{\mathcal U}\right)
	\qand \ddot \chi = 0
	~.
}
We choose to fix the $\U1$ current $J^0_{\U1} = \varrho\neq 0$ and $J_{\U1}^i = 0$, while we set the $\SU2$ currents completely to zero, $J_L^\mu=J_R^\mu=0$.
This corresponds (up to global $\U2$ transformations) to the following classical configuration:
\equ{
	\label{NLS:ClassicalEOMs}
	\chi = \mu\, t 
	\quads{,}
	\pi_1 = \frac{\pi}{4}
	\quads{,}
	\pi_2 = \pi_3 = 0
	~.
}
On the classical solution  
the $\U1$ current \eqref{NLSM:U2:U1Current} becomes
\equ{
	J_{\U1}^0 = \left(1 - \frac{c_{\mathcal R} \mathcal R}{2\mu^2}\right)F_0\, \mu^2
	\quads{\text{with}}
	F(1,1) \,\equiv\, \frac{F_0}{\sqrt2}
	~,
}
from where we deduce that $\mu \sim \order{\sqrt\rho}$, as expected. Hence, at large $\rho$ the chemical potential $\mu$
can be promoted to an expansion parameter for technical convenience.

\paragraph{Fluctuations}

Around the vacuum configuration described by \eqref{NLS:ClassicalEOMs} we parametrize the fluctuations by setting
\equ{
	\chi = \mu t + \frac{\hat \chi}{\sqrt\mu}
	\quads{,}
	\pi_1 = \frac\pi4 + \frac{\hat \pi_1}{\sqrt\mu}
	\quads{,}
	\pi_2 = \frac{\hat \pi_2}{\sqrt\mu}
	\quads{,}
	\pi_3 = \frac{\hat \pi_3}{\sqrt\mu}
	~.
} 
The normalization $1/\sqrt{\mu}$ is used so that the fluctuations have the proper dimension of a field (mass dimension $1/2$). 
The $\mu$-expansion of the dimensionless variables introduced in Eq.\  \eqref{NLS:DimensionlessVariables} around the classical ground state $\braket{\chi}$ and $\braket{\pi_i}$ gives 
\equ{
	X  
	= 1 + \frac{1}{\mu^3} \sum_{i=1}^3\left[\dot\pi_i^2-(\nabla\pi_i)^2\right]
	+ \order{\mu^{-7/2}}
	\quad,\quad
	Y
	= 1 + \frac{2}{\mu^3} \sum_{i=1}^3\dot\pi_i^2 + \order{\mu^{-7/2}}
	~.
}
As a consequence, $\mu$-expanding the Lagrangian up to order one we find
\begin{align}
\label{NonLinear:U2:FluctuatingLagrangian}
\mathcal L =&\,\, 
F_0 \left(\frac{\mu^3}{3}+ \frac{c_{\mathcal R}}{2} \mathcal R\, \mu \right) \,+\, F_0\, \mu^{3/2} \left(\partial_0\hat\chi\right)
\\
&
+\sfrac16\sum_{i=1}^3 \left[\left(3F_0+2F_X+4 F_Y\right)(\partial_0\hat\pi_i)^2 - \left(3 F_0+2F_X\right) \left(\nabla \hat\pi _i\right)^2 \right]
+F_0 \left[(\partial_0\hat\chi)^2 - \sfrac12 \left(\nabla \hat\chi\right)^2\right] + \order{\mu^{-1}}
~.
\nonumber
\end{align}
where 
\equ{
	\partial_X F(X,Y) \vert_{X=Y=1} \,\equiv\, \frac{F_X}{\sqrt2}
	\qand
	\partial_Y F(X,Y) \vert_{X=Y=1} \,\equiv\, \frac{F_Y}{\sqrt2}
	~.
}
Neglecting the total-derivative term, from the Lagrangian part which is quadratic in the fields it is straight-forward to determine the leading dispersion relations:
\equ{
	\label{NLS:OneRho:Dispersions}
	\omega_\chi = 	\frac{\abs{\textbf k}}{\sqrt2}
	~\qand~
	\omega_{\pi_i} = \sqrt{\frac{3 F_0+ 2 F_X}{3 F_0+ 2 F_X+4 F_Y}} \abs{\textbf k}~,\quad i=1,2,3
	~.
}
As expected from the analysis of the spectrum in the linear sigma model (section~\ref{ssc:TwoMus_OneU1}), we find four relativistic Goldstones, one with the universal dispersion relation ($c_s^{(1)}=1/\sqrt2$) and three with an undetermined, but equal, speed of sound $c_s^{(2)}$\,. 
Evidently, reality and causality, $c_s^{(2)}<1$, 
constrain the coefficients $F_0,F_X$ and $F_Y$ in the Taylor-expansion of $F(X,Y)$. 
Furthermore, the condensate contribution in the first line of Eq.\ \eqref{NonLinear:U2:FluctuatingLagrangian} entails two ignorance coefficients ($F_0$ and $c_{\mathcal R}$) in agreement with the general prediction of expansion \eqref{Intro:SchematicEnergyExpansion}.

Finally, note that in the non-linear sigma model for the present charge configuration the $\mu$-expansion
coincides with a field expansion in the Goldstones $\hat{\chi}$ and $\hat\pi_i$.
Expanding the  Lagrangian $\mathcal L$ up to order $Q^0$
results at most to a quadratic piece $\mathcal L^{(2)}(\hat{\chi},\hat{\pi_i})$.
Consequently, any term which is higher in the fluctuations $\hat{\pi}_i$, $\hat\chi$ will be automatically suppressed by powers of $1/Q$ and the derived dispersion relations \eqref{NLS:OneRho:Dispersions} are protected against quantum corrections.

\subsection{The $\U1\times\U1$ coset}

In a purely analogous fashion we can analyze the  general situation where $Q_1\neq Q_2$.
As we have seen in Section~\ref{ssc:TwoMus_TwoRhos} the low-energy spectrum is described by a  coset with $\U1\times\U1$ symmetry\footnote{A similar effective model has been constructed in \cite{Monin:2016jmo}, however not in the context of $\SU4$ matrix theory.}.
It is intuitive to parametrize the coset space via two Goldstone fields $\chi_1$, $\chi_2$ independently realizing each $U(1)$ symmetry as a shift,
$\chi_i \rightarrow \chi_i +\text{const}$ for $i=1,2$.
In the spirit of Eq.\ \eqref{NLSM:U2:BuildingBlock}
abbreviating 
\equ{
	\norm{\partial\chi} \equiv \sqrt{\abs{\partial\chi_1}^2 + \abs{\partial\chi_2}^2}
	\quads{\text{ where }}
	\abs{\partial\chi_i} \equiv \sqrt{\partial_\mu \chi_i\, \partial^\mu \chi_i}
	~,\quad
	i=1,2
	~,
}
the most general Lagrangian associated to the generic symmetry breaking pattern  can  be  written as
\equ{
	S = \int \d t \d\Sigma \left(\sfrac13\norm{\partial\chi}^3 - c_\mathcal{R} \mathcal R \norm{\partial\chi}\right) f \left(x, y\right)
	+
	\order{\mu^{-1}}
	~.
}
In this effective action, $f$ is an arbitrary function of the two dimensionless combinations, 
\equ{
	\label{NLSM:U1s:xy}
	x = \frac{\abs{\partial\chi_2}}{\abs{\partial\chi_1}}
	\qand
	y=\frac{\partial_\mu \chi_1\, \partial^\mu \chi_2}{\abs{\partial\chi_1}^2}
	~,
}
that can be independently constructed out of $\chi_1$ and $\chi_2$ in a Lorentz- and $\U1^2$\,-invariant way.
$\mathcal R$ is the Ricci scalar and $c_\mathcal{R}$ an arbitrary constant.
As in the previous example, this action makes sense only when the characteristic scale parameter $\mu$ (to be defined shortly) in $\norm{\partial\chi}\sim \mu$ is large.

\paragraph{The classical solution.}
Using the fact that the classical ground state of lowest energy at fixed charges is homogeneous in space, we solve the classical problem.
In detail, the  \textsc{eom}s read
\equ{
	\label{NLSM:U1U1:EOMs}
	\dot\chi_1 = \mu_1 = \mu \cos\ga
	\qand
	\dot{\chi}_2 = \mu_2 = \mu \sin \ga
	~,
}
introducing the time-independent chemical potentials $\mu_i$. It is convenient to parametrize them in polar coordinates in terms of the ``radial'' chemical potential $\mu$ and the angle $\ga$. Since this non-linear model should correspond to the generic charge configuration, also $\ga$ is taken to be generic.

For spatially homogeneous solutions there exists only one independent variable in Eq.\ \eqref{NLSM:U1s:xy}, $x=y$. 
On the classical $\textsc{eom}s$ it is $x=y=\tan\ga$.
In our homogeneous setting,
the two $\U1$ charge densities associated to the two independent shift symmetries in $\chi_i$ are given by 
\begin{align}
j_1^0 =&\,\, \frac{\delta \mathcal L}{\delta \dot\chi_1} 
= \left(\norm{\partial\chi} - \frac{c_{\mathcal R} \mathcal R}{\norm{\partial\chi}}\right) f(x,x)\, \dot\chi_1
-\left(\sfrac13\norm{\partial\chi}^3 - c_\mathcal{R} \mathcal R \norm{\partial\chi}\right) \partial_x f(x,x)\, \frac{\dot\chi_2}{\dot\chi_1^2} 
\quad \text{and}
\nonumber
\\[1ex]
j_2^0 =&\,\, \frac{\delta \mathcal L}{\delta \dot\chi_2} = 
\left(\norm{\partial\chi} - \frac{c_{\mathcal R} \mathcal R}{\norm{\partial\chi}}\right) f(x,x)\, \dot\chi_2
+\left(\sfrac13\norm{\partial\chi}^3 - c_\mathcal{R} \mathcal R \norm{\partial\chi}\right) \partial_x f(x,x)\, 
\frac{1}{\dot\chi_1}
~.
\end{align}
Abbreviating
\equ{
	f_0 \equiv f(\tan\ga,\tan\ga)
	\quads{,}
	f_x \equiv \left.\partial_x f(x,y)\right\vert_{x=y=\tan\ga}	
	\qand
	f_y \equiv \left.\partial_y f(x,y)\right\vert_{x=y=\tan\ga}
	~,
}
the conditions that fix two different charge scales at the vacuum described by Eq.\ \eqref{NLSM:U1U1:EOMs} can be combined as follows:
\equ{
	\label{NSLSM:U1s:FixedChargesCondition}
	\text{i=1,2\,:}\quad j_i^0 \overset{!}{=} \rho_i  
	\quads{\Rightarrow}
	\varrho^2 \equiv\rho_1^2 + \rho_2^2 = \mu^4 \left( f_0^2 +
	\frac{\left(f_x+f_y\right)^2}{9\cos^4\ga} + \order{\mu^{-2}}\right)
	~.
}
From here we  deduce the relevant scaling $\mu \sim \order{\sqrt{\varrho}}$\,, thus we can use $\mu$ as an expansion parameter at the technical level.

\paragraph{The fluctuations.}
We parametrize the fluctuations around the classical vacuum \eqref{NLSM:U1U1:EOMs} as 
\equ{
	\chi_1 = \mu t\cos\ga + 2\hat{\chi}_1/\sqrt{\mu}
	\quads{,}
	\chi_2 = \mu t \sin\ga + 2\hat{\chi_2}/\sqrt{\mu}
	~.
}
Dropping boundary terms, the Lagrangian  up to quadratic order in $\hat{\chi}_i$ (which coincides with $\order{1}$ in the large-$\mu$ expansion) reads
\begin{align}
&\mathcal L^{(0)} + \mathcal L^{(2)} =
f_0 \left(\mu^3  \, - c_\mathcal{R}\mathcal R\, \mu\right)
\\
+&\,\,
\left[3 f_0 (3+\cos2 \alpha)+4 \left((\cos\alpha)^{-2}-3\right)\left(f_x+f_y\right)\tan\alpha\right] (\partial_0\hat{\chi}_1)^2 
-\left[6 f_0-\frac{2 \left(f_x+2 f_y\right)\sin\alpha}{\cos^3\alpha}\right] \left(\nabla \hat\chi _1\right)^2 
\nonumber
\\
+&\,\,
\left[3 f_0(3-\cos 2 \alpha )+12 \left(f_x+f_y\right)\tan \alpha  \right](\partial_0\hat{\chi}_2)^2 
-
2 \left[3 f_0+2f_x \left(\sin2\alpha\right)^{-1}\right] \left(\nabla \hat\chi _2\right)^2 
\nonumber
\\
+&\,\,
\left[6 f_0 \sin 2 \alpha -8 \left(2\left(\cos\alpha\right)^{-2}-3\right) \left(f_x+f_y\right)\right] (\partial_0\hat{\chi}_1)(\partial_0\hat{\chi}_2)
-4 f_y\left(\cos\alpha\right)^{-2}\, \nabla\hat\chi_1\cdot\nabla\hat\chi_2
\nonumber
~,
\end{align}
Diagonalizing $\mathcal L^{(2)}$ we find for generic $\ga$ two relativistic Goldstones.
The one relativistic mode has the universal and the other a model-dependent
speed of sound:
\equ{
	c_s^{(1)} = 1/\sqrt2
	\qand
	c_s^{(2)} \equiv c_s^{(2)} (\ga,f_0,f_x,f_y)
	~.
}
The latter dispersion depends on the microscopic details of the underlying model through the ignorance function $f$.
Similar to the outcome in the linear sigma model, $c_s^{(2)} $ is independent of the large scale, 
while it depends on the ratio $Q_2/Q_1$, through $\ga$ here and $\vartheta$ in Eq.\ \eqref{LSM:GenericCase:SpeedsOfSound}. 
Finally, the suppression of the higher loops in the Goldstone fields follows 
immediately by derivative counting, as in the previous example.


\section{Conclusion and outlook}

In this work we have studied the universality class of three-dimensional theories with global $\SU N$ symmetry at the \ir fixed point
of the \rg flow where the order parameter can be described by a spin-0 field in the adjoint representation of the symmetry group. 
This unveiled a new aspect of \textsc{cft}s at large  charge, namely the possibility to fix multiple global $\U1$ scales while the classical ground state still remains homogeneous in space, in contrast to $O(2N)$ vector models.

In particular, we have seen that the low-energy spectrum of an $\SU N$ matrix theory at $\floor{N/2}$ fixed global charges is generically described by $\floor{N/2}$ relativistic Goldstone fields.
Similarly to previous studies,
by taking at least one of those global charges to be large we are able to perform perturbative calculations in the chosen sector of the strongly coupled theory.
The non-trivial prediction for the anomalous dimension of the lowest scalar operator with different charge assignments, given in  
Eq.~\eqref{LSM:OneRho:AnomalousDimension}, \eqref{LSM:GenericRhos:AnomalousDimension} and \eqref{LSM:TwoRhos:AnomalousDimension}
in the case of $\SU4$ matrix model, remains to be verified by non-perturbative methods, e.g.\ via simulations on the lattice.
%

Furthermore, the large-charge analysis enables us to make sharper statements about the structure of the space of theories with a given global symmetry.
So far, collecting what we know in the literature including the current paper, we see that the space of scalar theories with global $\SU N$ symmetry has at least  three  classes of qualitatively distinct fixed points:
\begin{itemize}[leftmargin=0.7cm]
	\item
	Wilson-Fisher fixed point with matter in the vector representation (i.e.\ the $O(2N)$ vector model)
	\item
	Qualitatively similar 
	to  Wilson-Fisher fixed point but with matter in the adjoint representation
	\item
	Qualitatively and quantitatively distinct fixed point with matter in the adjoint representation 
\end{itemize}
Since the various fixed points produce different predictions to tractable order, the way to probe them is to consider different charge configurations for the lowest scalar operator, cf.\ formula 
\eqref{Intro:SchematicEnergyExpansion} for $P=1,2$.
In fact, we do not even need to introduce multiple scales to see the novel character of the third fixed point.
Even when all $Q_j=Q$, this fixed point produces a distinct prediction ($f_2\neq0$ in Eq.~\eqref{Intro:SchematicEnergyExpansion}) compared to the first two.
{In particular, taking $Q_1=Q_2$ in our $\SU4$ application 
led to the emergence of a new symmetry breaking pattern in fixed-charge theories.
This symmetry breaking was independently analyzed using the non-linear effective description in Section~\ref{ssc:NonLinear_U2Coset}.}

The provided list of fixed points is by no means exhaustive for theories with global $\SU N$ symmetry. 
On the one hand, non-perturbative and beyond-the-leading-order effects  can further differentiate among the classified fixed points. 
On the other hand,
starting from $N=8$ the number of invariant terms one needs to write in the linear action of the matrix model to be sufficiently general to derive the dispersion relations becomes very large and hence the current effective techniques are no longer efficient. Still, the three distinct classes of fixed points outlined above remain inside special regions in the space of Wilsonian parameters, but more qualitatively different fixed points can arise.
It thus remains an open question how to efficiently tackle those models entailing significantly  larger symmetry groups. 
One could hope to relate that particular question to the more general investigations concerning the compatibility of large-charge and large-$N$ expansions {as well as to possible applications of large-charge techniques in the context of the \textsc{a}\small{d}\normalsize\textsc{s}/\cft correspondance.}

\pagebreak
\subsubsection*{Acknowledgements}

The author would  like to thank
Nezhla Aghaei, Luis Alvarez-Gaume, Costas Bachas, Dieter L\"ust, Riccardo Rattazzi, Cornelius Schmidt-Colinet and Uwe-Jens Wiese for providing useful insights to various aspects of the project.
The author is  grateful to
Domenico Orlando and Susanne Reffert for very helpful and extended discussions on the topic.
This work is supported by the Swiss National Science Foundation (\textsc{snf}) under grant number \textsc{pp}00\textsc{p}2\_157571/1.

\appendix 

\section{Tree level propagators}
\label{app:Propagators}

In this Appendix we explicitly provide the various blocks of the inverse propagators in momentum space used to semi-classically derive the dispersion relations in the linear sigma model.
\\

Specifically, in Section~\ref{ssc:TwoMus_OneU1} the  inverse propagator of the limiting model with $Q_1=Q_2$ was defined in Eq.\ \eqref{ToyModel:PropagatorBlocks} such that it takes block--diagonal form. 
The first block is given by the radial modes $r_1$\,,\,$r_2$  coupled  to the first two relativistic Goldstones along the Cartan directions $\Sigma^1$ and $\Sigma^2$\,:
\begin{align}
\label{LSM:OneRho:FirstPropagatorBlock}
\left.D^{-1}(k) \right\vert_{r_1,r_2,\chi_1,\chi_2}
=
\left(
\renewcommand*{\arraystretch}{1.5}
\begin{array}{cccc}
\textbf k^2-k_0^2 +M_r^2 & \frac{2(16 \lambda _1-3 \lambda _2)}{16 \lambda _1-\lambda _2} \mu ^2 & -2 i  k_0 \mu  & 0 
\\
\frac{2(16 \lambda _1-3 \lambda _2)}{16 \lambda _1-\lambda _2} \mu ^2 & \textbf k^2-k_0^2 +M_r^2 & 0 & -2 i k_0 \mu  
\\
2 i k_0 \mu  & 0 & \textbf k^2-k_0^2  & 0 
\\
0 & 2 i k_0 \mu  & 0 & \textbf k^2-k_0^2
\end{array}
\right)
\\[1ex]
\text{with}\quad 
M_r^2 = \frac{2\left(16 \lambda _1+\lambda _2\right)}{16 \lambda_1-\lambda_2} \mu ^2  +\frac{\mathcal{R}}{8}
~.
\nonumber
\end{align}
Notice that there is also a mixed term $r_1r_2$\,.
On the other hand, $r_\ga$ couples to $\chi_i$ for $\ga=i=3,4$ contributing two identical  blocks to the inverse propagator: 
\begin{align}
\left.D^{-1}(k)\right\vert_{\chi_3,r_3,\chi_4,r_4} = 
\left(
\begin{array}{cccc}
\textbf k^2-k_0^2 & 2 i k_0 \mu  & 0 & 0 
\\
-2 i k_0 \mu  & \textbf k^2-k_0^2 + m_r^2 & 0 & 0 
\\
0 & 0 & \textbf k^2-k_0^2 & 2 i k_0 \mu  
\\
0 & 0 & -2 i k_0 \mu  & \textbf k^2-k_0^2+ m_r^2
\end{array}
\right)
\renewcommand*{\arraystretch}{1}
\\[1ex]
\text{with}\quad 
m_r^2 = \frac{8 \gl_2}{16\gl_1-\gl_2}\mu^2 + \frac{\mathcal R}{8}
\nonumber
~.
\end{align}
The Goldstone--radial part in $D^{-1}(k)$ is followed by the diagonal contributions from the rest of the radial modes,
\equ{
	\left.D^{-1}(k)\right\vert_{r_5,r_6,r_7} = \diag \left(\underbrace{\textbf k^2 - k_0^2 + \frac{(16 \lambda _1 +7 \lambda _2)}{16 \lambda _1-\lambda _2} \mu ^2-\frac{\lambda _2 }{16 \lambda _1-\lambda _2}\,\mathcal{R}}_{3 \text{ times}}\right) 
	~,
}
together with those of the spectator fields in $U_\varphi$\,,
\equ{
	\left.D^{-1}(k)\right\vert_{\varphi_1,\varphi_2,\varphi_3,\varphi_4} = \diag \left(\underbrace{\textbf k^2 - k_0^2 + \mu^2}_{4 \text{ times}} \right) 
	~.
}
\\

In Section~\ref{ssc:TwoMus_TwoRhos} the block--diagonal inverse propagator was introduced in Eq.~\eqref{LSM:TwoMus:FullInversePropagtor} for a charge configuration with generic $Q_1\geq Q_2\geq0$, i.e.\ generic $\mu_1\geq\mu_2\geq0$ in the first Weyl chamber. 
We have the following blocks:

%
\begin{flalign}
\label{TwoMus:GoldstonePropagator}
\left.D^{-1}(k)\right\vert_{r_1,r_2,\chi_1,\chi_2,r_3}= &&
\end{flalign}
\vspace{-0.7cm}
\begin{flalign}	
\left(
\begin{array}{ccccc}
\textbf k^2-k_0^2 + m_{r_1}^2& \frac{v^4}{8}  \left(16 \lambda _1-3 \lambda _2\right)\sin \vartheta  & -2 i k_0 \mu _1 & 0 & 0
\\
\frac{v^4}{8}  \left(16 \lambda _1-3 \lambda _2\right)\sin \vartheta  & \textbf k^2-k_0^2 + m_{r_2}^2 & 0 & -2 i k_0 \mu _2 & 0
\\
2 i k_0 \mu _1 & 0 & \textbf k^2-k_0^2 & 0 & 0
\\
0 & 2 i k_0 \mu _2 & 0 & \textbf k^2-k_0^2 & 0
\\
0 & 0 & 0 & 0 & \textbf k^2-k_0^2 + m_{r_3}^2
\end{array}
\right)\
~,
\nonumber
\end{flalign}
\begin{align}
\text{with}\quad
m_{r_1}^2=&\,\,
\left(3 \lambda _1+\frac{3 \lambda _2}{8}+2 \lambda _1 \cos \vartheta +\frac{7}{8} \lambda _2 \cos \vartheta +\frac{5}{16} \lambda _2 \cos 2 \vartheta \right) v^4 +\frac{\mathcal{R}}{8} -\mu _1^2
~,
\nonumber
\\
m_{r_2}^2=&\,\,
\left(3 \lambda _1+\frac{3 \lambda _2}{8}-2 \lambda _1 \cos \vartheta -\frac{7}{8} \lambda _2 \cos \vartheta +\frac{5}{16} \lambda _2 \cos 2 \vartheta\right)v^4 + \frac{\mathcal{R}}{8} -\mu _2^2
~,
\nonumber
\\
m_{r_3}^2=&\,\,
\left(\frac{3}{2} \cos ^2\vartheta \,g_1 + \frac{9}{2} \cos^2 \vartheta\,g_3+\lambda_1+\frac{\lambda_2}{2}+\frac{1}{16} \lambda_2 \cos 2 \vartheta\right)v^4+\frac{\mathcal{R}}{8}
\nonumber
~,
\end{align}
which includes the Goldstone fields from $U_G$ coset factor together with the radial modes $r_\alpha$ in $\Phi_\text{radial}$\,.
%
In the same manner, $D^{-1}(k)\vert_{\varphi_i}$  
includes the spectator fields from $U_\varphi$ which are generally expected to be massive. According to \eqref{LSM:TwoMus:FullInversePropagtor} it breaks into a $4\times4$ part

\begin{flalign}
\label{TwoMus:SpectatorPropagator1}
\left.D^{-1} (k) \right\vert_{\varphi_i,\,i=1,...,4} = &&
\end{flalign}
\vspace{-0.6cm}
\begin{flalign}	
\renewcommand*{\arraystretch}{1.3}
\left(
\begin{array}{cccc}
-k_0^2 + \textbf k^2 + m_1^2 & \sfrac{1}{2} \sin\vartheta \left(\mu_1^2-\mu_2^2\right) & -i \cos\vartheta\,  k_0 \left(\mu _1+\mu_2\right) & -i \sin \vartheta\, k_0 \left(\mu _1+\mu_2\right)
\\
\sfrac{1}{2} \sin \vartheta  \left(\mu_1^2-\mu_2^2\right) & - k_0^2 + \textbf k^2 + m_4^2 & -i \sin \vartheta \,k_0 \left(\mu _1+\mu _2\right) & i \cos\vartheta\,  k_0 \left(\mu _1+\mu _2\right)
\\
i \cos\vartheta\, k_0 \left(\mu_1+\mu_2\right) & i \sin \vartheta\, k_0 \left(\mu_1+\mu_2\right) & - k_0^2 + \textbf k^2 + m_1^2 & \sfrac{1}{2} \sin \vartheta  \left(\mu_1^2-\mu_2^2\right)
\\
i \sin\vartheta\, k_0 \left(\mu_1+\mu_2\right) & -i \cos\vartheta\, k_0 \left(\mu_1+\mu_2\right) & \sfrac{1}{2} \sin\vartheta \left(\mu_1^2-\mu_2^2\right) &  - k_0^2 + \textbf k^2 + m_4^2
\end{array}
\right)
~,
\nonumber
\end{flalign}
together with another $4\times4$ block
\begin{flalign}
\label{TwoMus:SpectatorPropagator2}
\left.D^{-1} (k) \right\vert_{\varphi_i,\,i=5,...,8} = &&
\end{flalign}
\vspace{-0.6cm}
\begin{flalign}	
\renewcommand*{\arraystretch}{1.3}
\left(
\begin{array}{cccc}
-k_0^2 + \textbf k^2 + m_2^2 & \sfrac{1}{2} \sin\vartheta \left(\mu_1^2-\mu_2^2\right) & i \cos\vartheta\,  k_0 \left(\mu _1-\mu_2\right) & i \sin \vartheta\, k_0 \left(\mu_1-\mu_2\right)
\\
\sfrac{1}{2} \sin \vartheta  \left(\mu_1^2-\mu_2^2\right) & - k_0^2 + \textbf k^2 + m_3^2 & i \sin \vartheta \,k_0 \left(\mu_1-\mu_2\right) & i \cos\vartheta\,  k_0 \left(\mu_1-\mu_2\right)
\\
-i \cos\vartheta\, k_0 \left(\mu_1-\mu_2\right) & -i \sin \vartheta\, k_0 \left(\mu_1-\mu_2\right) & - k_0^2 + \textbf k^2 + m_2^2 & \sfrac{1}{2} \sin \vartheta  \left(\mu_1^2-\mu_2^2\right)
\\
-i \sin\vartheta\, k_0 \left(\mu_1-\mu_2\right) & i \cos\vartheta\, k_0 \left(\mu_1-\mu_2\right) & \sfrac{1}{2} \sin\vartheta \left(\mu_1^2-\mu_2^2\right) &  - k_0^2 + \textbf k^2 + m_3^2
\end{array}
\right)
~,
\nonumber
\end{flalign}
%
%
\begin{align}
\text{with}\quad
m_1^2 = \sfrac14\left(\mu _1-\mu _2\right)^2+\sfrac12 \cos \vartheta  \left(\mu _1^2-\mu _2^2\right)
&\qand
m_2^2 = \sfrac14\left(\mu _1+\mu _2\right)^2+\sfrac12 \cos \vartheta  \left(\mu _1^2-\mu _2^2\right)
~,
\\[1ex]
m_3^2 = \sfrac14\left(\mu _1+\mu _2\right)^2-\frac{(\cos 2 \vartheta -3) \left(\mu _1^2-\mu _2^2\right)}{4\cos \vartheta}
&\qand
m_4^2 = \sfrac14\left(\mu _1-\mu _2\right)^2-\frac{(\cos 2 \vartheta -3) \left(\mu _1^2-\mu _2^2\right)}{4\cos \vartheta }
~,
\nonumber
\end{align}
as well as two diagonal entries
\equ{
	D^{-1} (k) \vert_{\varphi_i,i=9,10}	=
	\diag \left(- k_0^2 + \textbf k^2 +\mu_1^2 ~,~ - k_0^2 + \textbf k^2  +\mu_2^2\right)
	~.
}


%
%

{\small	
\providecommand{\bysame}{\leavevmode\hbox to3em{\hrulefill}\thinspace}
\frenchspacing
\newcommand{\origttfamily}{}
\let\origttfamily=\ttfamily
\renewcommand{\ttfamily}{\origttfamily \hyphenchar\font=`\-}

}

\end{document}